\documentstyle[aps,preprint]{revtex}

\def\lambdabar{\mathchar'26\mkern-10mu\lambda }
\title{Binary Decay of Light Nuclear Systems}

\author{S.J. Sanders}

\address{University of  Kansas, Lawrence, KS 66045 USA}

\author{A. Szanto de Toledo}

\address{Departamento de F\'{i}sica Nuclear, 
Instituto de F\'{i}sica da Universidade
de S\~{a}o Paulo, 
Caixa Postal 66318-05315-970 - S\~{a}o Paulo, Brazil}

\author{C. Beck}

\address{\it Institut de Recherches Subatomiques, UMR7500, CNRS-IN2P3 et
Universit\'e Louis Pasteur, 23 rue du Loess, B.P.28, F-67037 Strasbourg, Cedex
2, France } 

\begin{document}
\maketitle

\newpage

\begin{abstract}
A review of  the characteristic features found in fully energy-damped, 
binarydecay yields from light heavy-ion reactions with $20\leq   A_{target} +
A_{projectile}\leq  80$ is presented.   The different aspects of these yields
that have been used to support models of compound-nucleus (CN) fission and
deep-inelastic dinucleus orbiting are highlighted.  Cross section calculations
based on the statistical phase space at different stages of the reaction are
presented and compared to the experimental results. Although the statistical
models are found to reproduce most of the observed experimental behaviors, an
additional reaction component corresponding to a heavy-ion resonance or
orbiting mechanism is also evident in certain systems.   The system dependence
of this second component is discussed. The extent to which the binary yields in
very light systems $(A_{CN} \leq 32)$ can be viewed as resulting from a
fusion-fission mechanism is explored. A number of unresolved questions, such as
whether the different observed behaviors reflect characteristically different
reaction times, are discussed. 
\end{abstract}

\section{Introduction}

As the interaction energy for a heavy-ion reaction increases above the Coulomb
barrier, the reaction grazing angle moves to more forward angles so that, by a
value of about 20\% above the barrier,  very small quasielastic and
single-nucleon transfer reaction yields are expected at larger scattering
angles.  Instead, the incident flux that might be expected to scatter to larger
angles, corresponding to smaller impact parameters,  is trapped by the
formation of a compound nucleus. For lighter systems, this compound system
subsequently decays by light particle and $\gamma$-ray emission,  with a very
small heavy-fragment $(A > 4)$ emission component. The experimental observation
of significant large-angle, elastic-scattering cross sections at energies well
above the Coulomb barrier has therefore been viewed with considerable interest.
Initially seen in systems of comparable target and projectile masses
\cite{ob75}, anomalous large-angle elastic cross sections have been found in a
number of systems. An early explanation for these yields \cite{ob75} was in
terms of a possible elastic-transfer mechanism where, simply, the target and
projectile were viewed to exchange identities. Alternatively,  optical model
calculations have also been found to predict enhanced back-angle yields when
weak absorption is assumed for  the grazing partial waves. Complicating the
picture is the observation that not all systems are found to demonstrate the
elastic scattering enhancements. Also,  some systems which exhibit the
phenomena involve the scattering of particles of very different masses. As a
further complication, many of the systems demonstrating a large-angle elastic
scattering enhancement are also found to show resonant behavior in scattering
and transfer reaction excitation functions. An overview of the anomalous
large-angle scattering (ALAS) phenomenon  is given in the review paper of
Braun-Munzinger and Barrette \cite{bmb82}. 

Although several mechanisms could be identified as possible reasons for
enhanced large angle yields in elastic and quasielastic channels, it was still
a surprise when Shapira {\it et al.} \cite{sfgsd79} discovered similar
large-angle yield enhancements for strongly energy-damped inelastic channels of
the $^{12}$C+ $^{28}$Si reaction. The energy spectra for these channels  were
found to be peaked at values that one would expect if the $^{12}$C and
$^{28}$Si fragments are emitted at rest from a sticking  configuration of the
compound system.  Further, angular distributions of constant
$d\sigma/d\theta_{c.m.}$ suggest a long-lived intermediate complex.  These
features led to the suggestion that an orbiting, dinucleus configuration is
formed that decays  back to the entrance channel. Weak absorption of the
grazing partial waves, as previously suggested in the elastic case, is
necessary to avoid having the orbiting configuration spreading into the
compound nucleus states. However, whereas for the elastic scattering
enhancements it was only necessary to invoke weak absorption for the grazing
partial waves, the orbiting phenomenon suggests weak absorption for even lower
partial waves with values between the critical angular momentum for fusion and
the reaction grazing angular momentum. 

After the discovery of orbiting in the $^{12}$C+$^{28}$Si system, similar
enhancements of large-angle, binary-reaction yields were subsequently observed
in somewhat heavier systems. While studying the  strong resonance behaviors
found in excitation functions of the $^{28}$Si+$^{28}$Si
\cite{bdp81,bbg81,bs83}  and $^{24}$Mg+$^{24}$Mg \cite{zkbsh83} elastic and
inelastic channels,  a significant non-resonant background yield was discovered
in the energy spectra of these channels at higher excitation energies.  This
yield was found to extend to larger angles. Other evidence for damped, binary
yields was found in a study by Grotowski {\it et al.} \cite{gmp84} of symmetric
mass fragments from the $^{12}$C+$^{40}$Ca, $^9$Be+$^{40}$Ca and
$^6$Li+$^{40}$Ca reactions. Using coincidence techniques, they established the
existence of significant decay strength to the mass-symmetric channels for
these systems. It is unlikely that this behavior could arise from a direct
reaction mechanism because of the large difference in mass asymmetry between
the entrance and exit channels. Angular distribution and energy spectra
measurements were shown to be consistent with the fission  decay of the
respective compound systems. Although these measurements were done at
relatively forward angles, the observed angular dependence would suggest
significant large-angle cross sections. 

Concurrent with the experimental measurements of energy-damped, binary reaction
yields in lighter systems came the development of finite-nuclear-range
corrections to macroscopic-energy calculations for these systems. By taking
better account of the role of the nuclear surface in determining nuclear
binding energies, these calculations made it possible for the first time to
obtain reasonable estimates of fission barrier heights in lighter systems
\cite{sierk86}. Whereas the standard rotating liquid drop model indicates high
fission barriers,  thereby precluding the prediction of strong fission
competition in these systems, the barrier energies found with  the new
calculations suggest that  significant fission competition might occur. 

The possibility of fission competition in light systems raises the question of
how one can distinguish between the  orbiting and fusion-fission mechanisms.
This is difficult to answer, although a clearer understanding of the respective
processes has been emerging through a number of different studies, as will be
discussed in this report. Some of the conceptual differences between the two
mechanisms are indicated schematically in Fig. 1.  Both processes are seen to
involve near-grazing  impact parameters with, however, the fusion-fission 
mechanism viewed as proceeding through the formation of a fully equilibrated
compound nucleus. The fission decay of the compound nucleus is determined by
the phase space available at a ``transition" configuration and, for light
systems,  can lead to significant population in many of the energetically
allowed mass channels.   For the orbiting mechanism, the system becomes trapped
in a more deformed configuration than that of the compound nucleus and is
inhibited from spreading into the compound nucleus states. This  allows for
significant decay back to the binary channels.  Although orbiting can be
considered as a deep-inelastic scattering mechanism, the rapid mass
equilibration that is thought to occur in light systems can result in the
population of channels with significantly different mass asymmetry than the
entrance channel. Still, it is expected that the orbiting mechanism will retain
a greater memory of the entrance channel than the fusion-fission process. 

The strong resonance behavior observed in excitation-function measurements of
large-angle elastic and inelastic scattering yields in several light systems
suggests that some structural aspect of the compound system may be strongly
influencing these yields.   This behavior is most pronounced in the
$^{24}$Mg+$^{24}$Mg system \cite{zkbsh83} where measurements \cite{qzkp90} have
indicated a resonance spin close to, or  exceeding, the grazing angular
momentum for the reaction. A possible explanation is that a highly deformed,
metastable configuration for the compound system is formed.  The relationship
between the resonance behavior and the fission process, the latter of which is
observed to occur at some level in all of the  reactions studied at energies
above the Coulomb barrier, is still an open question. A review of the resonance
aspects of heavy-ion breakup reactions,  including a discussion of earlier
electrofission measurements,   is found in the review article of Fulton  and
Rae \cite{fr90}. 

This paper will review the status of our  experimental and theoretical
understanding of energy-damped, binary-decay processes in light systems.
Although the relationship between these processes and that responsible for
heavy-ion resonance behavior will be explored, the general topic of heavy-ion
resonances in light systems  will not be discussed and  the reader is referred
to several review papers that already cover this general topic
\cite{fr90,eb85,gps95}. In Sec. \ref{sec:exp} we will explore the general
experimental signatures of the binary yields observed in lighter systems.   In
Sec. 3 the experimental techniques used to study these behaviors  will be
presented and some of the experimental difficulties highlighted. In Sec. 4 the
various models that have been developed to explain the  observed yields are
discussed. In Sec. 5 we will develop the  experimental systematics by
presenting a system-by-system discussion of current data, comparing these data
to the model calculations. In Sec. 6 we will present some of the open problems
and suggest possible measurements that might address these problems.  Finally,
in Sec. 7 we will summarize the present review. 

\section{Experimental Behaviors of Binary Decay Yields}
\label{sec:exp}
\setcounter{section}{2}
\setcounter{equation}{0}

One way to characterize nuclear reactions is in terms of the number of nucleons
exchanged between the incident particles.  Alternatively, one can consider the
degree to which the kinetic energy of the incident channel is transferred to
internal excitation of the outgoing fragments.  Clearly, there can be a
significant correlation between the number of nucleons transferred and the
energy damping that occurs in the reaction.  In our discussion of the general
reaction properties,  we use a somewhat arbitrary division that is motivated by
the experimental observables of mass, angle, and energy excitation of the
fragments.  We start with a discussion of the large-angle elastic scattering
yields.  We then consider the situation where one observes some energy damping,
and possibly mass flow,  but still do not have angular distributions of
constant $d\sigma/d\theta_{c.m.}$  that would signify a very long lived
complex.  Finally, we consider the situation where all degrees of freedom of
the system appear to have reached full equilibration. 

\subsection{Large-angle elastic and
quasi-elastic scattering}

After the first observations in heavy-ion reactions involving near-identical
particle systems \cite{ob75} of large-angle elastic-scattering yields in excess
of the optical model expectations, these backward-angle enhancements were found
to be a relatively common feature of  collisions between p-shell and sd-shell
nuclei \cite{bmb82}.  Figure 2 shows an angular distribution for the
$^{16}$O+$^{28}$Si reaction at $E_{lab}=55$ MeV.  The dashed curve is the 
prediction of an optical model calculation using the strong absorption E18
potential of ref. \cite{cdgz76}. The large angle yield is far in excess of the
model calculation and demonstrates a highly oscillatory behavior. As shown in
the insert, the angular distribution suggests a single orbital angular momentum
of $\ell=26\hbar$ may dominate the large-angle yields. An excitation function
of the backward angle yields for this system shows strong resonance-like
enhancements. This behavior might result from the formation of  nuclear
molecular configurations \cite{eb85,gps95,tmc82}, to an orbiting phenomenon
\cite{s82} related to the dynamics of the interaction potential, or to
compound-elastic scattering, that is, the fission decay of the compound nucleus
back to the elastic channel. Two or more of these mechanisms may coexist. 

The explanation for the resonance-like structures that are often found
associated with backward angle elastic and quasi-elastic yields in light
systems might be found in the interplay between interaction barriers and energy
dissipation mechanisms. The observed large-angle cross sections suggest the
formation of a long-lived nuclear molecule or dinucleus system that fragments
at a large deformation. This is consistent with the sticking model of deep
inelastic collision processes for which two colliding nuclei adhere to each 
other and undergo finite rotation before separating, with the possible
transformation of orbital angular momentum into spin of the outgoing fragments
and with a possible loss of kinetic energy of relative motion.  The resonances
might correspond to particularly simple configurations of the dinucleus
complex.  In a 1982 review of the ``anomalous" large-angle scattering
phenomenon (ALAS), however,  Braun-Munzinger and Barrette \cite{bmb82} have
shown the difficulty of attributing a specific reaction mechanism to the
resonance behavior. The current focus is to establish a global systematics for
the occurrence of ALAS behavior based on  the occurrence of weak absorption of
the near-grazing partial waves \cite{baa94}. 

\subsection{ Deep-inelastic
scattering yields}

Deep-inelastic reactions between heavy nuclei $(A_{target} +
A_{projectile}>50)$ typically involve substantial rearrangement of the nuclear
matter of the incident particles, with strong damping of the entrance-channel's
kinetic energy and transfer of orbital angular momenta to the spins of the
outgoing fragments \cite{v78,sh84}. The composite system may reach a
configuration close to the compound-nucleus saddle point before subsequent
elongation and fission. Wilczynski \cite{w73} has developed a qualitative
interpretation of deep-inelastic reactions in terms of a nuclear orbiting
process with dissipation of energy resulting from frictional forces. A
schematic illustration of the  progression from quasielastic scattering to
nuclear orbiting in heavy-ion collisions is given in Fig. 3 (adopted from ref.
\cite{w73}). The two  dimensional plot shows a contour of maximum cross section
(``Wilczynski  diagram") that contains the components of the three possible
classical trajectories, as drawn in the figure. Trajectories 1 and 2 correspond
to near-side and far-side scattering from the target nucleus.  In trajectory 3
the dinucleus system survives long enough to complete close to, or more than,
one full rotation. It is associated with an ``orbiting" process and leads to
emission spectra and emission probabilities that are independent of angle
\cite{sh84}.  That is, there is complete energy damping resulting in the final
kinetic energy of the fragments being close to the relative potential energy at
 the point where the fragments are closest together.  Also, the fragments'
angular distributions of $d\sigma/d\Omega$  follow a  $1/\sin \theta_{c.m.}$
angular dependence (corresponding to angular distributions of constant
$d\sigma/d\theta_{c.m.}$) suggesting the occurrence of a long-lived dinucleus
object in a sticking configuration. 

This early classical picture of the orbiting phenomenon has been developed
further using the classical models proposed, for example, by Bondorf {\it et
al.} \cite{bss74} and Gross and Kalinowski \cite{gk78}  for heavy-ion damped
reactions. 

The significant cross sections ($>10$ mb) observed for large-angle  inelastic
scattering yields of reactions such as $^{20}$Ne+$^{12}$C \cite{sfgsd79},
$^{27}$Al+ $^{16}$O \cite{sfg80}, $^{28}$Si+ $^{12}$C \cite{s82}, and
$^{24}$Mg+$^{12}$C \cite{gdbh90} were initially interpreted in terms of the
formation of a dinucleus, orbiting configuration.  These large yields were
found to be very surprising since any system in this mass range and making such
close contact, as suggested by an orbiting behavior, was expected to fuse into
a compound nucleus. The picture that emerges is one where the interacting ions
are trapped in a pocket of the ion-ion potential that results from the combined
effects of the Coulomb, centrifugal, and nuclear interactions. The orbiting
complex may  act as a ``doorway" to fusion, but weak absorption may also allow
decay back to the binary channels, possibly after significant mass flow and
rotation of the nuclear configuration. The description of an equilibrium 
orbiting model will be developed in Sec. \ref{sec:eqorb}. In subsequent
investigations it  has been shown that at least some of the yield attributed to
orbiting may correspond instead to the process of  fusion followed by fission. 
Still, experimental evidence for a distinct orbiting process in light systems
will be presented with the discussion of the $^{28}$Si+$^{12}$C and
$^{28}$Si+$^{14}$N  reactions. A non-compound origin is surmised for the
reaction yields in these systems based on their entrance-channel dependence
\cite{rk85}. 

\subsection{Fully energy-damped binary yields}

\subsubsection{Angular Distributions}

Enhanced large-angle yields in heavy-ion reactions reflect a relatively long
reaction time allowing significant rotation of the intermediate complex formed
in the reaction.  For the mechanism of fusion followed by fission, the evolving
intermediate complex is taken  to pass through a fully equilibrated compound
nucleus.  In the orbiting picture,  a long-lived dinucleus configuration is
envisioned.   In both cases the binary yields are assumed to correspond to
peripheral reactions involving incident angular momenta close to the critical
angular momentum for fusion $\ell_{cr}$. 

In considering the fission of heavier mass systems than those discussed here,
angular distribution data have yielded significant insight on the details of
the breakup mechanism \cite{sk97}. The ``reference" distributions for these
studies are those predicted by the standard transition-state model of fission
\cite{vh73}.  Using heavy-ion reactions to populate the systems,  it is
commonly found that the orbital angular momenta {\bf $L$} are large compared to
the projectile and target spins, resulting in the projection of the total
angular momentum {\bf $J$} on the space fixed z-axis, the beam direction, to be
small. Assuming  a Gaussian distribution of {\bf $K$}, the projection of  {\bf
$J$} on the symmetry axis of  the fissioning system,  and in the limit of zero
projectile and target spins, the expected fission angular distribution becomes
\cite{vh73} 
\begin{equation} \begin{array}{rll}
W(\theta)\propto &
\Sigma^\infty_{J=0}(2J+1)T_J  & \\
& \times 
{\displaystyle 
\Sigma^J_{K=-J}
{(2J+1)|d^J_{M=0,K}(\theta)|^2 \exp (-K^2 /2 K^2_0)\over
\sum\limits^J_{K=-J}\exp (-K^2/2K^2_0)}
} ,\\
\end{array}
\label{eqn:ang}
\end{equation}
where the $T_J$'s are the transmission coefficients giving the combined
probability for formation and fission decay of a compound nucleus with $J=L$
and $d^J_{MK}$ is the reduced rotation matrix.  The variance of the $K$
distribution can be expressed as \[K^2_0=I_{eff}T/\hbar^2 ,\] where \[{1\over
I_{eff}}={1\over I_\parallel}-{1\over I_\perp}\] and $I_\perp$ and
$I_\parallel$ are the moments of inertia about axes perpendicular and parallel
to the symmetry axis, respectively.  The ``transition'' state is typically
taken as the saddle point, defined as the configuration where the potential
energy of  the compound system as a function of deformation and mass asymmetry
is at its maximum. The temperature at the saddle point $T$ is found by
$T=\sqrt{E_x/a}$, where $E_x$ is the energy of internal excitation at the
saddle point and $a$ is the level density parameter.  Deviations between the
observed angular distributions and those predicted by the standard
transition-state model using the above expression for $W(\theta)$ have been
used as evidence for  faster processes than  that of complete fusion followed
by fission. 

Deep-inelastic scattering mechanisms,  where the characters of the incident
nuclei are maintained, can also lead to large angle yields if a dinucleus
configuration is reached that survives for a significant fraction of the
rotation period. If a large number of partial waves contribute to the reaction,
the resulting angular distributions become relatively featureless.  For
heavy-ion reactions, where the angular momenta associated with processes
leading to large angle yields tends to be high, the resulting angular
distributions approach the classical limit associated with systems emitting
fragments perpendicular to the rotation spin vector, with ${d\sigma /
d\theta_{c.m.}} = C$ or, equivalently, ${d\sigma / d\Omega}={C' / \sin
\theta_{c.m.}}$, where $C$ and $C'$ are angle-independent normalization
constants.  For the systems considered in this report, the differences between
the angular distributions predicted by the standard transition state model and
the ${d\sigma / d\Omega}={C' / \sin \theta_{c.m.}}$ distribution that
characterizes a classical, long-lived orbiting configuration are only evident
at very forward  and backward angles. This is illustrated for the
$^{35}$Cl+$^{12}$C reaction at $E_{lab}= 180$ MeV \cite{bdhd89,bdhf92}. Angular
distributions of fission-like fragments are shown in Fig. 4 for the inclusive
yields to different mass channels. One finds distributions of constant
${d\sigma / d\theta_{c.m.}}$ for each of the measured isotopes. In Fig. 5 the
angular distribution data in ${d\sigma / d\Omega}$ for the Ne channel are
compared to both a  $1/\sin \theta_{c.m.}$ angular dependence and the angular
dependence predicted by the transition state model (using Eqn. \ref{eqn:ang}).
For this latter calculation the transmission coefficients for the fission
yields $T_J$ were based on the transition-state calculation  to be discussed
later in this report and the level density parameter was taken as $a =
A_{CN}/8$.  The two models show overlapping behavior in the  region of the
experimental data and significant differences between the calculations are only
observed at very forward and backward angles. Unfortunately, the cross sections
at angles near to 0$^\circ$ and 180$^\circ$ are also very difficult to measure.
 Large elastic scattering yields can obscure the forward angle data and low
particle energies can make the large angles yields difficult to detect.   In
general it has not been possible in lighter systems to deduce information about
the time scales of the more strongly damped processes based on angular
distribution data. 

A more rapid rise in the energy-inclusive cross sections at forward angles than
given by a 1/sin$\theta_{c.m.}$ distribution indicates a shorter lifetime of
the composite system, with breakup occurring within the first revolution of the
system.  Such lifetimes are incompatible with the formation of an equilibrated
compound nucleus, but may still reflect significant energy damping within a
deep-inelastic mechanism.  By measuring the forward peaked distributions, it is
possible to estimate the lifetime of the intermediate nuclear complex using a
diffractive, Regge-pole model. The distributions are fitted by 
\begin{equation}
{d\sigma\over d\Omega}={C\over \sin \theta_{c.m.}}
\left\{ {{\rm e}^{-
\theta_{c.m.}/\omega t} +{\rm e}^{-(2\pi
- \theta_{c.m.})/\omega t}}\right\} .
\label{eqn:life}
\end{equation}
This expression describes the decay of a rotating dinucleus with angular
velocity $\omega ={\hbar \ell / \mu R^2}$ where $\mu$ represents the reduced
mass of the system, $\ell$ its angular momentum (which should fall somewhere
between the  grazing $\ell_g$ and critical $\ell_{cr}$ angular momentum), and
$R$ represents the distance between the  two centers of the dinucleus. Small
values of the ``life angle" $\alpha(\equiv \omega\tau$) lead to forward peaked
angular distributions associated with fast processes, whereas large values of 
$\alpha$, associated with longer times as compared to the dinucleus rotation
period $\tau$, are consequently associated with longer lived configurations and
lead to more isotropic angular distributions. In the limiting case of a very
long-lived configuration, the distributions approach a $d\sigma / d\Omega
\propto 1 / \sin \theta_{c.m.}$ dependence. 

One way to characterize the sensitivity of the angular distributions to the
``life angle" $\alpha$ is by considering the anisotropy function $R(\theta)$,
where \[R(\theta)={d\sigma\over d\Omega}(\theta,\alpha)/ {d\sigma\over
d\Omega}(90^o,\alpha)  . \] The functional dependence of $R(\theta)$  with
$\alpha$ is shown in Fig. 6 for $\theta = 10^o$ and $170^o$.  The corresponding
anisotropy for an angular distribution with ${d\sigma / d\Omega}\propto {1 /
\sin \theta_{c.m.}}$ is shown by the dashed line. The plot suggests that  an
anisotropy  measurement might be sensitive to the time  scale of fast processes
$ (\alpha < 90^o)$, but is less sensitive to the time scale of strongly damped
processes with $\alpha > 120^o$. 

Measurements of excitation energy integrated yields of specific mass channels
show, in some systems, angular distributions which reflect the contributions of
both short- and long-lived processes.  As  an example, the observed
distributions for inclusive, energy-summed yields for the  $^{16}$O+$^{11}$B
reaction at $E_{lab}=64$ MeV are shown in Fig. 7 \cite{aacf94}. It should be
noted that for this particular reaction, the isotope pairs C-N, B-O, and F-Be
correspond to specific exit channels.  From this figure one finds a 
contribution of forward peaked processes (N and F fragments) with
$\alpha<70^o$, corresponding to direct mass transfer,  with slower processes
(C, B, and Be fragments) presenting large ``life angles", with $\alpha \gg
180^o$. 

\subsubsection{Energy Spectra}

For fully energy-damped reactions, the energy spectra are peaked at large
negative $Q$-values, with the cross section maxima typically occurring at a
value of total kinetic energy in the exit channel that corresponds to the
relative potential energy of the two nascent fragments in a near-touching
configuration. This observation is consistent with the idea that the composite
system passes  through a stationary point in the potential energy surface where
the relative  kinetic energy is at a minimum.  The stationary point could be
the saddle point of a compound nucleus undergoing fission or, alternatively,
the dinucleus  configuration of an orbiting binary complex. The total kinetic
energy of the  fragments (TKE) in the exit channel is then expected to  equal
the sum of the nuclear and Coulomb potential energies and the rotational energy
of the  rotating complex at the stationary point, with \[TKE = V_{nucl}(d) +
V_{Coul}(d) + V_{rot}(d) ,\] and $V_{nucl}$, $V_{coul}$, and $V_{rot}$ are the
nuclear, Coulomb, and relative rotational energy between the two nascent
breakup fragments at a separation $d$ corresponding to the stationary point. In
general, the sticking limit is assumed in calculating the rotational energy
contribution. 

The total kinetic energy of the outgoing fragments is therefore sensitive to
the deformation of the system at the time of scission
\cite{s82,e76,bmcg76,cbmc77,nnegg77,a80,sgmd80,rdg85}. In general, the most
probable TKE values are well described by assuming stationary shapes consistent
with the predicted  saddle-point configurations of the nuclear systems
\cite{s91}.   The exceptions to this rule tend to occur for nuclear systems
where anomalous large-angle elastic scattering (ALAS) cross sections are
observed. 

It should be noted that even though large kinetic energy damping is one of the
experimental signatures used to distinguish a damped reaction process from
direct reactions in which only a small part of the initially  available kinetic
energy is dissipated, it is not necessarily the case that a given
fusion-fission or orbiting reaction event will have a large  energy loss.   The
distribution of the $TKE$ values about the peak value can be large and, in some
cases, may lead to the population of the ground states of the outgoing
fragments, as would be the case for compound elastic scattering. 

The dependence of the average energy release in fission on atomic number and
mass of the compound nucleus has been widely studied for heavy systems. A
linear relationship between the most probable TKE release value with
$Z^2/A^{1/3}$ of the fissioning nucleus has been established by the systematic
work of Viola {\it et al.} \cite{vkw85}, with this systematics more  recently
extended to the lighter systems in ref.\cite{tt92}.  The compiled $TKE$ values
\cite{bmetc95} of the symmetric fission fragments produced in light heavy-ion
systems are shown in Fig. 8. The original Viola systematics \cite{vkw85} 
(dashed line) are capable of describing the whole data set except in the case
of low $Z$ fissioning nuclei. In these very light systems,  the diffuse nature
of the nuclear surface and the associated perturbations of the necking degree
of freedom, as calculated using the finite-range, liquid-drop model
\cite{ajs85}, results in a change in the predicted slope, leading to vanishing
$TKE$ values as $Z$ approaches zero.  This effect is observed experimentally
and is reproduced by the solid  curve in Fig. 8, which is calculated  using
\cite{tt92}: \[TKE = Z^2/(aA^{1/3}+bA^{-1/3}+cA^{-1})\] where the values of the
fitting parameters are a=9.39 MeV$^{-1}$, b=-58.6 MeV$^{-1}$ and c=226
MeV$^{-1}$, respectively. 

\section{Experimental Arrangements}
\setcounter{section}{3}
\setcounter{equation}{0}

Several considerations influence the  experimental techniques used to study the
processes considered in this report. In a given measurement most of the
processes that compete with those being studied correspond to more peripheral
collisions and result in yields at angles near to or forward of the grazing
angle. This requires that angular distributions be extended to larger angles
where the more peripheral mechanisms have little yield, although with the
possible experimental complication of the energy of the reaction fragments
falling below the identification thresholds of the detectors  being used. 

To provide a complete description of the reaction process, it becomes necessary
to determine,  for each fragment, the mass(A), atomic number (Z), energy ($E$)
(or velocity ($v$)),  and the emission angle ($\theta$) of the particle.  In
cases where the  fragments are emitted in binary decays,  the conservation laws
for mass, energy, and momentum can be used to reduce the number of quantities
needed to be measured to fully specify the reaction.  If one or both of the
fragments is at an energy above its particle emission threshold, however, there
is likely to be a secondary light particle emitted before the  fragment is
detected. The effects of such secondary light particle emission need to be
considered. 

In the simplest arrangements for detecting single particles,  element
identification (Z) and energy measurement ($E$) is generally accomplished  by
determining differential energy loss  $\Delta E$ and the total energy $E$  in a
telescope consisting of a gas $\Delta E$ counter backed by a solid-state
detector.  The $\Delta E$ signal is proportional to the square of the charge. 
By stopping the particle in the solid state detector, the  residual energy is
also determined and thus the total energy can be deduced.   Alternatively, a
Bragg-curve detector can supply the information on $Z$ and $E$ \cite{ku94}.  In
either case, the desired cross sections tend to be low (typically,
$d\sigma/d\Omega < $5 mb/sr), requiring large solid-angle detectors and,
possibly, necessitating the  use of position-sensitive detectors for angle
determination. 

For measurements done at pulsed beam facilities it is possible to determine the
energy and mass of a particle through time-of-flight and total energy
measurement.  In this case a single, bare Si(surface-barrier) detector can be
used to measure the yields.  The detection of  low energy particles may still
be  difficult, however, as a consequence of multiple scattering effects of the
particle leaving the detector and the dependence of the timing signal on the
energy and charge of the particle incident on the detector. In general, 
depending on the preferred detection technique, experiments have either
measured the charge or mass of the outgoing fragments, but not both
simultaneously. 

As an alternative to the singles measurement of the reaction products, for a
binary decay it is also possible to determine the mass and energy of the
outgoing fragments by measuring the scattering angles of both fragments and the
relative timing of the two fragments.  This technique has proven to be quite
valuable since it allows for the use of inexpensive, large-area,
position-sensitive avalanche detectors (PPACs).  Figure 9 illustrates the basic
variables of the measurement.  The two outgoing particles are detected at
laboratory scattering angles of $\theta_1$ and $\theta_2$, respectively, in
detectors located at distances of $d_1$ and $d_2$ from the target. The relative
time of arrival at the detectors, $\Delta t=t_2-t_1$, can be measured directly
using a time-to-amplitude circuit, or can be deduced from measurements of the
individual flight times. In the non-relativistic limit, the linear momenta of
the two particles in the laboratory system, $\vec p_1$ and $\vec p_2$, can be
determined from the total laboratory momentum $\vec p_o$ and the scattering
angles, with 
\[
p_1=p_o{\sin (\theta_2)\over \sin (\theta_1+\theta_2)}, \;
{\rm and}
\]

\[
p_2=p_o{\sin (\theta_1)\over \sin (\theta_1+\theta_2)}
.\]
The masses of the two scattered particles can be related to the above
quantities  and the projectile and target masses, $A_{proj}$ and $A_{targ}$,
with \[A_2={(d_1
{\displaystyle A_{proj}+A_{targ}\over 
{\displaystyle p_1}}+\Delta t
)\over 
{\displaystyle
({d_2\over p_2}+
{d_1\over p_1}) 
}},\;     {\rm and}\]

\[A_1=A_{proj} + A_{targ} - A_2.\]
It is then possible to determine the energies of the two particles and the
corresponding reaction $Q$-value using the derived masses, with 
\[E_1 = E_{lab}{A_{proj}\over A_1} [{\sin (\theta_2)\over
\sin (\theta_1+\theta_1)}]^2\; ,\]

\[E_2=E_{lab}{A_{proj}\over A_2} [{\sin (\theta_1)\over
\sin (\theta_1+\theta_1)}]^2\; ,\; {\rm and}\]

\[Q=E_1+E_2-E_{lab}\; ,\]
where $E_{lab}$ is the laboratory beam energy. If single mass resolution is
achieved (the typical situation for the lighter systems being considered), then
the reaction $Q$-value can be subsequently determined by using only the
position information and the known masses. In this case, the excellent position
resolution that can be achieved with the PPAC's can result in $Q$-value
resolutions of $<200$ keV, with the limiting resolution determined largely by
multiple-scattering effects as the particles leave the target. 

The coincidence detection of both reaction fragments can also be useful in
cases where one or both of the fragments is  formed with sufficient excitation
energy to result in a secondary emission of a  light particle (typically an
$\alpha$ particle). On average, the emitted light particles result in a 
distribution of fragment velocities about their original, pre-evaporation
values.  In this case, the deduced mass distribution reflects the
pre-evaporation values.  If, in addition, the fragment mass is measured through
an alternative method, one can determine the extent to which the secondary
evaporation modifies the observed mass distribution from the original,
preevaporation distribution \cite{setc89}. 

An essential element of most model calculations of the binary decay yields is
an estimate of the spin distribution of the compound nucleus, as usually
determined through measurement of the total fusion cross section. In general,
the experimental determination of the fusion cross section is obtained by
measuring the evaporation residues yields and, in the case of a fusion-fission
analysis, adding in the fission yields.   When pulsed beams are available, the
evaporation-residue yields can be  measured using bare, Si(surface barrier)
detectors.  Otherwise, it is possible to employ time-of-flight telescopes using
 micro-channel-plate detectors (MCP), or to use a set of $E$-$\Delta$E
telescopes. 

Figure 10 shows the experimental arrangement used in a study of the binary
yields from the $^{24}$Mg+ $^{24}$Mg reaction \cite{hsfp94} that has many of
the elements discussed above.  The Bragg-curve detector \cite{ku94} determined
the charge and energy of outgoing fragments.  A large multigrid avalanche
counter (MGAC 2) \cite{wplf87} determined the angles of the particles entering
the Bragg-curve detector.  A second avalanche counter (MGAC 1) was used to
detect the coincident fragments and determine their angles.  Finally, a number
of Si(surface barrier) detectors were used for normalization purposes and to
obtain a measure of the evaporation-residue cross section. 

The use of angular distribution data to separate different reaction components
requires that these distributions be measured over as large an angular range as
possible.  For a given fragment, energy thresholds set by the detectors may
severely restrict the largest angle that can be reasonably measured. In the
case of a binary decay, however, it may be possible to use a measurement of the
corresponding recoiling fragment to extend  the angular distribution. Figure 11
demonstrates this simple concept for the  $^{18}$O+ $^{10}$B reaction going to
the $^{15}$N+ $^{13}$C exit channel. The ``beamlike" particle ($^{15}$N) going
to forward center-of-mass angles is detected at forward laboratory angles, as
indicated in the velocity diagram. Although ``reverse kinematics"  also results
in forward scattering angles when the $^{15}$N particle undergoes large
center-of-mass deflection, the laboratory energy of the particle is now small,
complicating particle identification.  At this point, however, the
``targetlike"  $^{13}$C particle is scattered to forward angles and can be
easily detected. Combining the measurements for the $^{15}$N and $^{13}$C
results in a more complete angular distribution for this channel.   Although
this procedure is only valid in cases where secondary light-fragment emission
can be ignored, it has been found to be very useful in the analysis of lighter
systems \cite{aacf94,atst93}. 

In studies where the interplay between the nuclear structure and reaction
dynamics is investigated,  one important technique has been to identify the
$\gamma$-decay cascade from the excited final fragments. In this case, the low
reaction cross sections mandate experimental arrangements with both large
particle and $\gamma$-ray detection  efficiencies.  The identified particles
are used for channel selection and Doppler correction of the $\gamma$-ray
spectra.  Such measurements have been done in studies of the $^{32}$S+
$^{24}$Mg \cite{shp94} and $^{36}$Ar+ $^{12}$C \cite{fsd96} reactions. 

\section{Model calculations}
\setcounter{section}{4}
\setcounter{equation}{0}

\subsection{ Statistical Models}

It took only a short time after the experimental observation of neutron-induced
fission for a quantitative description of the process to be developed in terms
of the transition-state model \cite{bw39}. In this model, where fission
competes with $\gamma$-ray and light-particle emission in the de-excitation of
the compound nucleus, the probability for fission is determined by the most
restrictive phase space (level density) encountered by the fissioning system
between the equilibrated compound-nucleus configuration and the exit channel. 
This  ``transition state" is usually taken as the configuration where the
macroscopic potential energy reaches its maximum value, the saddle point,
although discrepancies observed in heavier systems with the angular dependence
described by Eqn. 2.1  have also led to the consideration of the more deformed
scission point as the transition state \cite{pdb84,rhs84}. 

In lighter systems, where the saddle- and scission-point configurations are
expected to be very close and little damping is expected as the system proceeds
between the two, there is less reason to expect significant differences between
calculations done at the saddle and scission points than for heavier systems
where the shapes of the saddle- and scission-point configurations are quite
different and significant damping can occur in moving between the two. Indeed,
calculations based on saddle-point \cite{s91} and scission-point \cite{mbnm97}
transition-state configurations are found to give equivalent results in lighter
systems.  In this section the transition-state formalism will be reviewed and
the saddle-point  \cite{s91} and scission-point \cite{mbnm97} calculations
compared. 

The development of saddle-point calculations in lighter systems was
significantly delayed from the comparable development for heavy-system fission 
because of the difficulty in accounting for the finite range and diffuse
nuclear surface effects that strongly influence the macroscopic energies of
these systems.  In light nuclei, the saddle-point shapes correspond to two
deformed spheroids separated by a well-developed neck region, with the surfaces
of the two spheroids coming within close proximity of one another. This results
in strong surface effects. The development of the finite-range model
\cite{sierk86}, however, has made it possible to extend the saddle-point
calculations to very light systems.  Within this model, saddle-point energies
are found by determining the stationary points of the compound nucleus
potential-energy surface as a function of spin and constrained mass asymmetry. 
The shape parameterization, discussed by Nix in Ref. \cite{jrn68}, consists of
three connected quadratic surfaces of revolution. The calculations explicitly
account for the diffuse nature of the nuclear surface and the finite range of
the nuclear interaction. 

Since the development of the finite-range model, one remaining difficulty in
applying its results to the fission of lighter systems has been the lengthy
process of performing  individual calculations for all of the necessary spin
and mass-asymmetry  dependent saddle-points that might be encountered as the
system breaks apart. Whereas for heavier systems the mass-asymmetry dependence
of the macroscopic  energy favors symmetric breakup, in lighter systems an
asymmetric breakup is favored. This problem has been somewhat ameliorated with
the development of a simple, double-spheroid parameterization of these energies
\cite{s91} . In this double-spheroid picture, the saddle-point energy is 
given by \[ V_{saddle}(J_{CN}, \eta) = V_C + V_n + V_r + V_o ,\] where $J_{CN}$
is the spin of the compound nucleus, $\eta (= 1-2A_L/A_{CN})$ is the mass
asymmetry, $V_C$ is the Coulomb energy between two deformed spheroids, $V_n$ is
the nuclear  interaction energy given by the two-body, finite-range-model
potential of Krappe, Nix, and Sierk \cite{kns79}, and $V_r$ is the total
rotational energy of the  double-spheroid configuration, using a
diffuse-surface corrected moment of inertia.  $A_L$ is the atomic mass of the
lighter of the two fission fragments. The final term, $V_o$, accounts for the
influence of the saddle-point  neck as well as other, shape-independent aspects
of the macroscopic energy. The spheroid geometry is adjusted to reproduce the
full finite-range-model calculations of Sierk \cite{sierk86}. 

A comparison of saddle-point energies found using the double-spheroid model and
energies obtained from full macroscopic energy calculations is shown in Fig. 12
for several systems. Figure 13 (from Ref. \cite{s91}) compares some typical
saddle-points shapes obtained with the full calculations (solid curves) and the
corresponding double-spheroid shapes (dashed curves). The double-spheroid
parameterization is found to closely reproduce the saddle-point energies found
with the full calculations.  Moreover, the shapes obtained with the
double-spheroid  model are similar (excluding the neck) to the corresponding
finite-range-model configurations. Assuming that the saddle- and scission-point
configurations are similar, this suggests that the relative energy of the two
spheroids calculated within this model should be related to the total kinetic
energy $E^{tot}_K$ of the fragments in the exit channel. 

A schematic diagram of the energy balance for a fusion-fission reaction is
shown in Fig. 14, taken from Ref. \cite{shp94}. Starting from the entrance
channel with incident center-of-mass energy $E_{c.m.}$, the Coulomb and
centrifugal energy dominated entrance barrier must first be overcome to form a
compound nucleus with excitation energy $E^*_{CN}$ and spin $J$. The effective
excitation of the compound nucleus $E^*_{eff}$, which enters in calculating the
density of compound nucleus states, is expected to be somewhat less than
$E^*_{CN}$ (by an amount $\Delta_{eff}$) since, at these high energies, it is
assumed that the virtual ground states corresponds to the macroscopic-energy
ground state \cite{fp77}.  The most restricted region of phase space, where the
density of states becomes a minimum, occurs at the saddle point.  In Fig. 14,
$\varepsilon$ is taken as the kinetic energy associated with the radial motion
at the saddle point. This term further reduces the density of saddle-point
states.  A saddle-point shell correction $\Delta V_{shell}$ is also taken to
influence the fission decay probabilities.  The excitation energy available at
the saddle point $u_J$, which determines the corresponding level density
$\rho_f$, is then given by \[u_j=E^*_{CN}-V_{saddle}(J,\eta)- \Delta
V_{shell}-\Delta_{eff}-\varepsilon \] 

The probability for the compound nucleus of spin $J$ to break up to a fission
channel of mass asymmetry $\eta$ is proportional to the level density  $\rho_f$
 above the corresponding spin $J$ saddle point. Some authors have taken the
scission point as the  transition state, in which case the corresponding energy
of the two fragments at scission (not shown) is used to determine the
transition-state level density. It will be shown that the very small energy
difference $\delta$ believed to exist between the saddle and scission points
leads to very similar fission predictions of the saddle- and scission-point
models. Assuming $\delta \approx 0$, the total kinetic energy in the exit
channel, and, correspondingly, the reaction $Q$-value, can then be related to
the relative energy of the fragments at the saddle/scission configuration.   In
this case, taking $\varepsilon=0$ for the most probable decay probability
(corresponding to the highest level density), \[E^{tot}_K=V_C+V_n+{\hbar^2\over
2\Im_{rel}}\ell (\ell +1)\] with \[\ell={\Im_{rel}\over \Im_{tot}} J_{CN}.\]
The relative moment of inertia of the two spheroids is given by $\Im_{rel}=\mu
r^2$ and, taking the moments of inertia of the individual deformed spheroids as
$\Im_1$ and $\Im_2$, respectively, the total moment of inertia for the two
spheroid configuration is then given by \[\Im_{tot}=\Im_1+\Im_2+\Im_{rel}.\] 

Calculation of fission cross sections in the statistical models is based on the
Hauser-Feshbach formalism.  For a compound nucleus of spin $J$ that is
populated with a partial fusion cross section of $\sigma_J$, the partial
fission cross section is given in terms of the ratio of the fission decay width
$\Gamma^{fis}_J$ to the total decay width for this spin $\Gamma^{tot}_J$, with
\[\sigma^{fis}_J={\Gamma^{fis}_J\over \Gamma_J^{tot}} \sigma_J\; .\] The fusion
partial cross section for formation of a compound nucleus of spin $J$ from
projectile and target nuclei of spins $J_p$ and $J_t$, respectively, at
center-of-mass energy $E_{c.m.}$ is given by 

\[\sigma_J=\pi\lambdabar^2{2J+1\over
(2J_p+1)(2J_t+1)}\sum\limits^{J_p+J_t}_{S=|J_p-J_t|}
\sum\limits^{J+S}_{\ell=|J-S|} T_\ell (E_{c.m.})\; ,\]
with \[\sigma^{tot}_{fus}=\sum\limits^\infty_{J=0}\sigma_J.\] A simple and
commonly used method of representing the fusion transmission coefficient is to
take 
\[T_\ell(E_{c.m.}) = {1\over 1+exp \{[\ell -\ell_{cr}(E_{c.m.})]/\Delta\}}  
,\]
where $\ell_{cr}$ is the critical angular momentum for fusion and $\Delta$ is
the diffuseness of the fusion $\ell$-distribution. The critical angular
momentum for fusion $\ell_{cr}$ can either be obtained from fusion model
calculations or by adjusting its value to achieve consistency with measured
evaporation-residue cross sections.  Most of the lighter system calculations
have been done using $\Delta=1\hbar$. 

For the calculation of the total decay width $\Gamma_{tot}$, it is assumed that
the deexcitation of the compound nucleus is through the emission of neutrons,
protons, alpha particles, $\gamma$ rays and/or 
fission fragments.  Then
\[
\Gamma_{tot}=\Gamma_n +
\Gamma_p +\Gamma_\alpha+
\Gamma_\gamma + \Gamma_{fis}
.\]
Both the saddle-point calculations of Sanders \cite{s91} and the recent
scission-point calculations of Matsuse {\it et al.} \cite{mbnm97} use the code
CASCADE \cite{fp77} for calculating the partial widths for the three light
particles and $\gamma$ rays.  Here, the partial width $\Gamma_x$ for particle 
$x (x = n, p,$ or $\alpha$) of spin $s_x$ to be emitted from the compound
nucleus  of excitation energy $E^*_{CN}$ and spin $J_{CN}$ to form an
evaporation-residue nucleus ER of excitation energy $E^*_{ER}$ and spin
$J_{ER}$  is given by 
\[
\Gamma_x=\int{\rho_{ER}(E^*_{ER} -\Delta_{eff},J_{ER})\over
2\pi \rho_{CN}(E^*_{CN}-\Delta_{eff}, J_{CN})}
\sum\limits^{J_{ER}+S_x}_{S=|J_{ER}-S_x|}
\sum\limits^{J_{CN}+S}_{\ell=|J_{CN}-S|} T^x_\ell 
(\varepsilon_x)d\varepsilon_x
.\]
The integral is over all kinetic energies of the emitted light particle
$\varepsilon_x$, and $\rho_{CN}$ and $\rho_{ER}$ are the level densities of the
compound nucleus and resulting evaporation residue, respectively.  The
transmission coefficients $T_\ell^x(\varepsilon_x)$ are obtained from
optical-model calculations using average parameters. For the higher excitation
energies involved in the fission process, it is expected that shell effects
should have little influence on the level densities and hence an effective
macroscopic energy ground state is used to calculate these densities
\cite{fp77}.   The parameter $\Delta_{eff}$ determines the zero point of the
effective excitation energy (see Fig. 14), with
\[\Delta_{eff}(MeV)=E_B(Z,A)-E^{macro}_B(Z,A).\] Here $E_B$ is the measured
binding energy of the nucleus and $E^{macro}_B$ is the corresponding
macroscopic energy \cite{s91,fp77}. The partial width for $\gamma$ decay
assumes decay through the giant dipole resonance. 

\subsubsection{Saddle-point model}

In lighter systems, the mass asymmetry dependence of the fission barrier favors
the decay into mass asymmetric exit channels. This can be seen in a plot of the
calculated saddle-point energies $V_{saddle}$ as a function of the fragment
mass for the $^{56}$Ni compound system in Fig. 15. In this figure,
$A_{fragment}=28$ corresponds to symmetric breakup into two $^{28}$Si
fragments.  Typical saddle-point shapes are shown for spin $\ell =0\hbar$  at
two different mass asymmetries and for the symmetric barrier at $\ell =
36\hbar$. At low spins the very steep rise in the saddle-point barrier energy
in going to more symmetric configurations leads to a strong favoring of
light-fragment evaporation over heavy-fragment fission of the compound nucleus.
It is only at higher spin values that the more symmetric breakup channels
become able to compete in the compound-system decay. In developing a model for
the fission decay of light systems,  it  therefore becomes important to
consider the decay widths to specific exit channels, corresponding to different
mass asymmetries at the saddle point, with the total width given by the sum 
over the widths to the individual channels:
\[\Gamma_f = \sum\limits_{A_L}\sum\limits_{Z_L} 
\Gamma_f(Z_L, A_L).\]
Here the partial widths are denoted by the charge $Z_L$ and mass $A_L$ of the
lighter fragment. The corresponding mass asymmetry $\eta$  is given by
$\eta=1-2(A_L/A_{CN})$. 

In the saddle-point model \cite{s91} the fission widths $\Gamma(Z_L,A_L)$ are
obtained by considering the level density above the mass-asymmetry dependent
saddle point, with 
\[\Gamma_f(Z_L,A_L) = \int\limits^{20 MeV}_{\epsilon=0}
{\rho_f(u_J)\over
2\pi\rho_{CN}(E^*_{CN}-\Delta_{eff}, J_{CN})}
T^f_{J_{CN}}(\varepsilon)d\varepsilon ,\]
and where the transmission coefficients have a sharp cutoff form, with 
\[ 
\begin{array}{rll}
T^f_{J_{CN}}&(\varepsilon)=& \\
&\left\{ \begin{array}{ll}
1\; {\rm for}\; \varepsilon 
\leq E^*_{CN}-V_{saddle}(J_{CN},\eta)-\Delta
V_{shell}(J_{CN},Z_L, A_L)-\Delta_{eff} \\
0\; {\rm for}\; \varepsilon > 
E^*_{CN}-V_{saddle}(J_{CN},\eta)-\Delta
V_{shell}(J_{CN},Z_L, A_L)-\Delta_{eff} & \\
\end{array}
\right .
\end{array}
.\]

The integration is over the energy of radial motion $\varepsilon$  and is
insensitive to the upper limit assuming that this limit is sufficiently large.
$V_{saddle}(J_{CN},\eta)$ is the spin- and mass-asymmetry dependent
saddle-point energy with respect to the macroscopic-energy ground state of the
compound nucleus. The importance of the compound nucleus spin in determining
the competition between light particle evaporation and fission to heavier
fragments is illustrated in Fig. 16 where the partial  cross section
distribution for fusion (solid line) and fission (shaded region), as defined by
binary decay to a channel where the lighter fragment has mass $A_L\geq 6$, are
shown for the $^{32}$S+ $^{24}$Mg reaction at $E_{lab}$=121 MeV.  In this
system, fission is found to only compete at incident partial waves close to the
critical angular momentum for fusion. 

Although shell effects are not expected to influence the level density of the
equilibrated compound nucleus, at the much ``colder" saddle-point configuration
these effects can be important.  This is evidenced by a strong isotopic
dependence for the fission cross sections that is inconsistent with a smooth
dependence of the potential-energy surface on the mass-asymmetry parameter
$\eta$.  As a first approximation of the shell corrections at the saddle point,
a term $\Delta V_{shell}(Z_L,A_L)$ has been added to the barrier energies based
on the sum of the Wigner energy corrections \cite{mn81} for the two nascent
fragments: 
\[\Delta V_{shell}(Z_L,A_L)= W(Z,A_L) + 
W(Z_{CN}-Z_L, A_{CN} -A_L)\]
with
\[
W(Z,A)=(36 {\rm MeV}) \left[ |{A- 2Z\over A}| + 
\left\{ {1/A, Z\; {\rm and}\; N\;
{\rm odd\; and\; equal}\atop 0,\; {\rm otherwise}}  \right]
\right\} \]
This correction to the potential energy surface has the effect of enhancing the
fission cross section to channels where both fragments have  N=Z. 

To calculate the level densities of the compound nucleus and saddle-point
configurations, a Fermi-gas formula \cite{bm81} is used, with 
\[ \rho(u,J)={2J+1\over 12}\sqrt{a_x} 
\left[{\hbar^2\over 
2\Im}\right]^{3/2}
u^{-2}\exp (2\sqrt{a_x u})\]
and
\[ u= \left\{ \begin{array}{ll}
E^*_{ER} - {\hbar^2\over 2\Im}J (J+1)-\Delta, & 
{\rm ER} \\
E^*_{CN}-V(J_{CN},\eta)-
\Delta V_{shell}(Z_L, A_L)-\Delta_{eff} - \varepsilon, &
{\rm saddle\; point}
\end{array}
\right
. \]
For the evaporation residues (ER), the level-density parameter $a_x =a_n$ and
the spin $J$ and energy offset $\Delta$ correspond to the evaporation residue. 
  The saddle-point densities are calculated with $a_x=a_f$ and $J=J_{CN}$. Most
of the calculations done for light systems have used $a_n=A_{ER}/(8 {\rm MeV})$
and $a_f=A_{CN}/(8 {\rm MeV})$. 

In light nuclear systems, the fission breakup of the compound nucleus is often
to final  fragments where the density of states is low.  This can result in
considerable structure in the excitation spectra of the fission fragments.  To
explore the population of states in the final fragments while retaining the
saddle-point as the ``transition" state, a procedure has been developed
\cite{shp94} to calculate the population of specific mutual excitations
assuming a stochastic process.  Within the transition-state method, population
of a given saddle-point level already corresponds to a commitment to fission
into a particular mass partition. The partial cross section for the  population
of the compound nucleus with spin $J$ that subsequently undergoes fission to
mass asymmetry $\eta$ is taken as $\sigma_{FF}(J,\eta)$ and is calculated using
the transition-state model based on the saddle-point phase space.  The cross
section for populating a specific mutual excitation ($\beta_1,\beta_2)$ is then
given by 
\[ \sigma(\beta_1,\beta_2)=
\sum\limits_J\sigma_{FF}(J,\eta)
{ \sum\limits_{\ell_{out}}
[\beta_1\times \beta_2]_{J,\ell_{out}}
P(\eta,J, \varepsilon) \over
\sum\limits_{\lambda_1,\lambda_2,\ell_{out}}
[\lambda_1\times
\lambda_2]_{J,\ell_{out}} P(\eta, J, \varepsilon)}
,\]
where $[\lambda_1\times \lambda_2]_{J,\ell_{out}}$ represents the sum of the
possible spin couplings between the two fragments in states $\lambda_1$ and
$\lambda_2$ with orbital angular momentum $\ell_{out}$ and coupling to
compound-nucleus spin J, and $P(\eta,J,\varepsilon )$ is the probability of the
compound nucleus of spin $J$ to fission with mass asymmetry $\eta$ and radial
kinetic energy $\varepsilon$.  This probability depends implicitly on
$\ell_{out}$ through $\varepsilon$. 

The radial kinetic energy $\varepsilon$   can be expressed in terms of the
characteristic energies of the reaction with 
\[ \varepsilon = E_{c.m.} + Q_o-V_{rel}(\ell_{out}, 
\eta)+\delta - E_x.\]
The significance of each of these energies is shown schematically in Fig. 14.
Here, $E_{c.m.}$ is the center-of-mass energy in the entrance channel, $Q_o$ is
the ground-state $Q$ value, $V_{rel}(\ell_{out}, \eta)$ is the relative energy
of the two spheroids that comprise the saddle-point shape, $\delta$ is the
energy loss that occurs in moving from the saddle to scission configurations,
and $E_x$ is the mutual excitation of the final fragments.  In light systems
$\delta$ is expected to be small. 

Figure 17 from ref. \cite{shp94} compares the results of this calculation to
the observed excitation spectra for the $^{24}$Mg($^{32}$S,$^{28}$Si) $^{28}$Si
reaction at $E_{c.m.}=51.0$ and 54.5 MeV.  The bold-line histograms show the
experimental results. The dotted line histograms show the predicted spectra
using all known levels in $^{28}$Si up to the 14.339 MeV excitation.  The
thin-line histograms are the predicted spectra for only the particle-bound
levels. Since the experimental results were obtained using a kinematic
coincidence technique that discriminates against excitations where one or both
of the populated states subsequently emits a light particle, the thin-line
histograms are expected to more faithfully represent the experimental
situation. The observed structure is very well reproduced by the calculations.
The structure that is observed at higher excitation energies can be attributed
to groupings of mutual excitations with high channel spins. The essential
validity of the predicted population pattern was confirmed by a measurement of
the $\gamma$-rays in coincidence with the fission fragments for this system
\cite{shp94}, as will be discussed in Sec. \ref{sec:exp44to56}. 

One of the interesting features of the energy spectrum calculation is that it
suggests that higher excitation energies correspond to smaller compound nucleus
spins.  The calculation also suggests significant alignment of the spin and
orbital angular momenta at higher excitations, with the possibility of
anti-aligned configurations being prevalent at lower excitation energies.  This
is shown in Fig. 18 for the calculation shown in Fig. 17.  The average values
of the compound nucleus spin, orbital angular momentum, and channel spin is
shown for each mutual excitation of the $^{28}$Si+$^{28}$Si fission channel. 

\subsubsection {Scission-point model}

In the Extended Hauser-Feshbach, scission-point model, the partial fission
decay width is determined by the product of the level densities in the final
nuclei, with 
\begin{eqnarray*}
\Gamma_f(Z_L,A_L) & = &
 \sum\limits_{(I_L,I_H)I}\; 
\sum\limits_{(L,I)J}\int\int\int
{\rho_{I_L}(E^*_L)\rho_{I_H}(E^*_H)\over 
2\pi\rho_{CN}(E^*_{CN},J_{CN})} \\
& & \times \delta
(E^*_L + E_H^* + \varepsilon + Q - E^*_{CN})T_L 
(\varepsilon)d E^*_L d
E^*_H d\varepsilon 
\end{eqnarray*}
and where $E^*_L$ and $E_H^*$ are the excitation energies in the lighter and
heavier fragments, respectively,  $I_L$ and $I_H$ are the spins of the
fragments,  $\varepsilon$ is the relative energy of the fragments in the exit
channel, and  $Q$ is the $Q$-value for the binary breakup.  The delta function
assures energy conservation.  For low-lying excitations of  the fragments, the
integrals are replaced by summations over the discrete energy levels. 

In the development of the Extended Hauser-Feshbach model, various assumptions
have been made for the level densities and transmission coefficients needed to
evaluate the above expression.  In what is perhaps the most systematic study of
this model as it applies to light systems, the level density expression is
taken as given above for the saddle-point model and with level density
parameters given by $a=A/(8 {\rm MeV})$ \cite{mbnm97}. The  transmission
coefficients for this study are evaluated by using the simplified formula 
\[ T_L(E)={1\over 1+\exp \left\{ \left[ V(L)-E\right] /\Delta_s \right\} },\]
where the diffuseness parameter $\Delta_s$ is typically taken to equal 0.5 MeV.
 It has further been found possible to obtain good agreement with experiment
using a simple parametric expression of the barrier  height $V(L)$ at the
scission point, with 
\[
V(L) = V_{Coul}+{\hbar^2 \over 2\mu_f R^2_S}\ell (\ell +1)
,\]
where $\mu_f$ is the reduced mass of the decaying complex fragments.  The
scission point $R_S$ is estimated by using the radii $R_L=r_s A^{1/3}_L$ and
$R_H=r_s A^{1/3}_H$ of two spherical fragments of mass number $A_L$ and $A_H$
separated by a distance d, with \[R_S=R_L+R_H +d.\] The ``neck" parameter d for
the reactions covered by this report is found to vary from 2.5 to 3.5 fm using
$r_s$ = 1.2 fm. The Coulomb energy is calculated using the expression \[
V_{Coul}=Z_LZ_He^2/R_S ,\] where $Z_L$ and $Z_H$ are the atomic numbers of the
lighter and heavier fragments, respectively.  In general, it has been found
possible to reproduce the experimental results by varying the neck parameter
$d$ in a  systematic manner, leading to predicted fission cross sections
comparable to those obtained with the saddle-point model.  With this
adjustment, the moments of inertia for the two calculations are also similar,
supporting the assertion that, in lighter nuclear systems, the scission- and
saddle -point configurations are similar. 

Mass distribution calculations based on the saddle-point \cite{s91} and
scission-point models \cite{vkw85,bdhf93}, summing the fission and
evaporation-residue components for a given mass channel,  are compared to
experimental cross sections obtained  for the $^{35}$Cl+$^{12}$C reaction at
$E_{lab} = 200$ MeV and for the $^{23}$Na+$^{24}$Mg reaction at $E_{lab}$ =
89.1 MeV \cite{bdhf92,bdhf93} in Fig. 19. These two reactions reach the common
$^{47}$V  compound systems at approximately the same excitation energy of
$E^*_{CN}$= 64 MeV. The observed and calculated values for the average total
kinetic energies for the two reactions are shown in  Fig. 20, although here the
extended Hauser-Feshbach results for the $^{23}$Na+ $^{24}$Mg reaction are not
available.  Both calculations are found to give comparably good agreement with
the experimental results.  In both cases, the overall
fission+evaporation-residue cross sections observed experimentally were used to
determine the $\ell_{cr}$ values for the two systems.  Otherwise, in the
saddle-point calculations, all of the remaining parameters are set by the
general systematics of fission in light systems.  In the case of the extended
Hauser-Feshbach calculations, the neck separation parameter $d$ was adjusted to
optimize the fit to the fission data. However, as previously indicated, this
parameter is found to vary uniformly as a function of the mass of the compound
system. 

The saddle-point calculations have been extended to even lighter systems, most
notably by the S\~ao Paulo group exploring nuclei in the mass A=28 region
\cite{aacf94}, using the double-spheroid model to extrapolate saddle-point
energies to these systems.  Again, good agreement is achieved between the 
calculations and experimental results. 

\subsection{Equilibrium orbiting model}
\label{sec:eqorb}

The binary reaction yields for a number of light systems, including those for
the $^{24}$Mg+ $^{12}$C \cite{gdbh90}, $^{28}$Si+ $^{12}$C \cite{s82,ssf84},
$^{24}$Mg+ $^{16}$O \cite{kbg92}, and $^{28}$Si+ $^{14}$N \cite{sah87,ssa88}
reactions, have been interpreted in terms of the formation and decay of
long-lived, rotating, dinucleus complexes. A quantitative model for this
behavior has been developed by Shivakumar, Ayik and collaborators
\cite{ssa88,ass88} based on nucleon transport theory. However, because of the
long lifetime of the orbiting complex, as indicated by the angular distribution
data,  the equilibrium limit of the more general transport model is used in
comparisons with data.  In this limit, the model is somewhat similar to the
saddle-point model of fission, although the ``equilibrium orbiting" model
provides a more unified description of the fusion and deep-inelastic orbiting
processes. 

At energies near the Coulomb barrier, the collision of two heavy ions is 
governed by a barrier potential comprised of nuclear, Coulomb and centrifugal
terms.  The equilibrium orbiting model is based on the observation that at low
energies, and for all partial-waves up to some maximum value $\ell_{pocket}$,
the interaction potential exhibits a pocket as a function of the distance
coordinates. This pocket  allows for trapping of the incident particles and 
the subsequent fusion or orbiting behavior of the composite system. For even
higher angular momenta with $\ell_{pocket}\leq \ell \leq \ell_{max}$, trapping
still occurs as frictional forces can reduce the relative energy and angular
momentum to values of $\ell\leq \ell_{pocket}$.  The value of the maximum
angular momentum $\ell_{max}$ is set by when the centrifugal energy reaches a
value that prevents the incident particles from closing to within a critical
distance where the nuclear surfaces overlap. 

The trapped, dinucleus complex can either evolve with complete amalgamation
into a fully equilibrated compound nucleus or, alternatively, escape into a
binary exit channel by way of orbiting trajectories.  Orbiting can therefore be
described in terms of the formation of a long-lived dinucleus complex which
acts as a ``doorway" state to fusion with a strong memory of the entrance
channel. The observed orbiting yields correspond to the fragmentation of the
dinucleus complex during the early stages of the interaction, while long-lived
complexes  relax towards the mono-nuclear shape of the compound nucleus. 

A full description of the exchange of mass and charge between the two
interacting  ions requires solving coupled transport equations describing the
mass flow. The equilibrium model \cite{ass88} is an approximation to the full
transport theory that assumes the  probability of fragmentation into a channel
$\chi\equiv (N,Z)$ and angular momentum $\ell$ can be expressed as the product
of two terms. The first term is a distribution function that is based on the
density of states in $\chi$ available to the incident channel, normalized with
respect to the density of states for all channels, including the fusion
channel. The second term is a transitional probability that is assumed to be
slowly varying with energy and is thus taken as a constant. The defining
potential energy surface for the dinucleus complex is written as 
\[ \begin{array}{rll}
U_J(N,Z;R) = &V_n(N,Z;R)+V_C(N,Z;R)& \\
&+
{\displaystyle 
{\hbar^2\over 2{\Im}_{tot}(N,Z;R)}
}J(J+1)+Q(N,Z),& 
\end{array}\]
where $V_n$ is the nuclear interaction (taken as the empirical proximity
potential of Bass\cite{rb77}), $V_C$ is the Coulomb interaction,
$\Im_{tot}(N,Z;R)$ is the total system moment of inertia in the sticking limit,
$J$ is the total angular momentum, and $Q(N,Z)$ is the ground state $Q$-value
of channel $(N,Z)$ with respect to the entrance channel.  For long-lived
systems the decay distribution function $P_\ell(N,Z)$ is obtained from the
ratio of the density of states evaluated at the saddle point (conditional
saddle) of the potential, for a given $\ell$-value and exit channel, to the sum
over all open channels of the state densities at the respective saddle points
and the corresponding state densities evaluated at the configurations
corresponding to the minimum of the $\ell$-dependent potential-energy for each
open channel. The  total binary fragmentation probability is given by summing
over the open channels, with \[P_\ell=\sum\limits_{N,Z}P_\ell (N,Z)\] and the
corresponding fusion probability is then (1-$P_\ell$). It is therefore possible
to evaluate both the binary fragment and fusion probabilities within a
consistent formalism. The total production of a fragment (N,Z) is given by the
sum of all $\ell$-partial wave contributions up to the maximum angular momentum
$\ell_{max}$ for which the system can become trapped in a pocket of the
interaction potential: 
\[\sigma_\ell (N,Z)={\pi\over 
k^2}\sum\limits^{\ell_{max}}_{\ell = 0}(2\ell
+1)P_\ell (N,Z)\]
A corresponding expression gives the fusion cross section:
\[\sigma_{fus}={\pi\over 
k^2}\sum\limits^{\ell_{max}}_{\ell = 0}(2\ell
+1)(1-P_\ell)\]
The final total kinetic energy  is determined by the sum of the nuclear,
Coulomb, and rotational energies at the conditional saddle point, with: 
\[E^{tot}_{K,\ell}(N,Z)=U_o(N,Z;R_S)+ {\hbar^2\ell (\ell 
+1)f^2\over 2
\Im_{rel}(N,Z;R_S)}-Q(N,Z)  ,\]
where $f = \Im_{rel}(N,Z;R_S) / \Im_{tot}(N,Z;R_S)$ and
 $\Im_{rel}$ and $\Im_{tot}$ are the relative and total moments of inertia at
the saddle point configuration with radius $R_s$, respectively. The nuclear and
Coulomb contributions to the potential energy $U_o(N,Z;R_s)$ are evaluated at
$\ell=0$. The average kinetic energy in the exit-channel, taking into account
the channel cross section, is found using: 
\[<TKE(N,Z)>={\pi\over 
k^2}\sum\limits^{\ell_{max}}_{\ell=0} (2\ell
+1)E^{tot}_{K,\ell} {P_\ell(N,Z)\over P(N,Z)} ,\]
with
\[
P(N,Z)={\pi\over k^2}\sum\limits^{\ell_{max}}_{\ell=0} 
(2\ell+1)P_\ell(N,Z)
.\]

The equilibrium orbiting model has been used to successfully explain both the
observed cross sections and TKE values of the fully damped fragments for
several  lighter nuclear systems. It has been found that the adjustment of a
single strength parameter in the Bass parameterization  \cite{rb77}  of the
nuclear potential leads to good  agreement for both orbiting and total fusion
yields. This is shown in Figs. 21 and 22 \cite{sah87} for the
$^{28}$Si+$^{12}$C reaction \cite{s82,ssf84}. In Fig. 21 the measured total
kinetic energies for the carbon, nitrogen, and oxygen channels are compared to
the experimental results. Figure 22 compares the measured and calculated cross
sections for the same three channels as well as for the total
evaporation-residue cross section. In both of these figures the orbiting
calculations are indicated by the dotted curves. 

For comparison, the results of the transition-state, saddle-point calculation
discussed earlier are also shown in Figs. 21 and 22 by the solid curves. The
parameters used for the transition-state calculations are the same as those
used to successfully describe the experimental cross sections in other light
systems.  For the $^{28}$Si+ $^{12}$C reaction, however, the transition-state
calculations  tend to underestimate the observed binary yields.  This is seen
in the figures where the total fusion cross section has been adjusted to
achieve the best general agreement of the total kinetic energies and binary
reaction cross sections with experiment.  This leads, however,  to
fission-model predictions of evaporation residue cross sections which are high 
compared to the experimental results and predicted carbon cross sections that
are too small. 

In other reactions of very light systems, such as for the $^{28}$Si+ $^{14}$N
reaction, the equilibrium orbiting and transition-state models result in
comparable agreement with the data. For heavier systems ($A_{CN}\geq 47$), the
simple form of the equilibrium orbiting model is found to be less successful
\cite{bdhf92,setc89}. This may reflect, however, some of the simplifying
assumptions of the calculations, such as the calculation of moments of inertia
based on touching,  spherical fragments and the use of ground-state $Q$-values
for the driving potential energy surface.    To achieve a satisfactory
agreement for the $^{32}$S+ $^{24}$Mg reaction within the equilibrium orbiting
model \cite{setc89}, for example,  it has been found necessary to introduce
deformation effects by assuming a moment of inertia more similar to that of the
nuclear saddle point. In this case the equilibrium orbiting and scission-point
models employ very similar phase-space arguments. 

The shortcomings of the equilibrium model for orbiting does not imply that the
presence of an orbiting mechanism, as distinct from fission, can be ruled out
in some systems. The alternative statistical models also fail to explain
several important features of the orbiting reaction data, such as the strong
entrance channel dependence in the $^{40}$Ca system \cite{kbg92} and spin
distributions and alignments measurements for the $^{28}$Si+ $^{12}$C  reaction
\cite{rsh91}. 

\subsection{NOC}

The relationships among the different reaction mechanisms such as 
fusion-fission, orbiting, and heavy-ion resonances are still not fully
understood.  It has been suggested \cite{rrb84}, for example,  that the
heavy-ion resonance phenomenon observed in some light nuclear systems  may
reflect  a strong shell correction to the fission potential energy  surface
and, as such,  be related to the statistical fission mechanism. In  the absence
of a fully consistent picture relating the different reaction mechanisms,
considerable effort has gone into developing the systematics of  these
processes.  In general, the fully energy-damped yields are observed with a
cross section comparable to the predictions of the transition-state
calculations.   The occurrence of a distinctly different orbiting or resonance
component to the reaction cross section seems to be correlated with the number
of open reaction channels (NOC). In lighter systems, for example, the even-even
C+C, C+O, and O+O reactions all show strong resonance behavior and also have
small NOC values.  Underlying this correlation may be a  relationship between
the NOC for a given reaction and the ``surface transparency" of the reaction
\cite{ha81}. 

The NOC calculations have been systematically extended to heavier systems 
\cite{baa94} and a strong correlation found between the anomalous  large-angle
resonant structure and small NOC values.  For example, systems  populating the 
$^{40}$Ca compound nucleus, for which there is significant evidence of an
orbiting process \cite{kbg92}, have low values for the NOC corresponding to the
grazing partial waves.  Alternatively, the energy-damped yields for several
systems populating  the $^{47}$V system \cite{bdhd89,bdhf92,bdhf93}, where the
NOC values are large,  are consistent with a fusion followed by fission
picture. 

The NOC for a given system is obtained by a triple summation over all possible
two-body mass partitions in the exit channels, over all possible angular
momentum couplings and, finally, over the allowed excitations of the two
fragments: 
\[N^J(E_{c.m.})=\sum\limits_{A_1+A_2=A_{cn}}\; 
\sum\limits_{J = L+I_1+I_2}\;
\sum\limits_{E_r=E^*_{CN}-E_1-E_2-Q_{12}}T_L(E_r)
.\]
In this expression,  $E_{c.m.}$ and $E_{CN}^*$ are the center-of-mass energies
of the incident particle and the excitation energy of the compound system,
respectively. $A_1, A_2, A_{CN}$ are the mass numbers of the outgoing fragments
and the compound system. $I_1, I_2$ and $L$ are the intrinsic spins of the
fragments and the orbital angular momentum of their relative motion. $Q_{12}$
is the reaction $Q$-value of the decay into the fragments.  $E_1, E_2$ and
$E_r$ are the intrinsic excitation energies of the fragments and the energy
available to their relative motion, respectively. $T_L(E_r)$ is the
transmission coefficient of the outgoing channel as a function of angular
momentum and relative energy. The transmission coefficients have been
calculated using the semi-classical, parabolic barrier penetration
approximation: 
\[T_L(E_r)=1/\left[1+\exp\left[{2\pi(E_L-E_r)\over
\hbar\omega_L}\right]\right]\]
where $E_L=V_L(R_B)$.  The rotational frequency,
\[\hbar\omega_L=\hbar\sqrt{ \left[ ({d^2V_L(R)\over 
dR^2})|_{R=R_B}\right] / \mu }
,\]
is related to the curvature of the outer barrier. In this expression, $\mu$ is
the reduced mass and $V_L(R)$ is the sum of the Coulomb, centrifugal and
nuclear potentials.  In more recent calculations a macroscopic proximity form
has been used for the nuclear interaction \cite{baa94}, rather than the
Saxon-Woods form employed in earlier calculations.  The sum over the energy
sharing between the two fragments employs discrete energy levels of the
fragments at lower energies, where these are known, and an angular momentum
dependent level density expression at higher energies \cite{baa94}. The
sensitivity of the NOC calculations to the choice of level density expression
and to the transmission coefficients is discussed in Ref. \cite{baa94}. 
Although the quantitative results are sensitive to these choices, the
qualitative behavior is not, thus preserving the predictive value of the
calculations. 

The expression for $N^J$ is similar to the denominator appearing in the
Hauser-Feshbach formalism \cite{hf52} for the compound nucleus.  However, the
NOC calculations include direct reaction channels in addition to the
evaporation channels.  This allows the phase space calculation to be extended
to values of the incident angular momentum greater than the critical angular
momentum for fusion $L_{gr}$. 

In order to compare different systems, it is useful to normalize $N^J$ by the
corresponding incident flux.  The quantity N/F is defined, with 
\[N/F=N^J(E_{c.m.})/F^J(E_{c.m.})\;
.\]
Here, $F^J(E_{c.m.})$ is the incident flux for the total angular momentum J, as
given by 
\[F^J(E_{c.m.})={\pi\over 
k^2}\sum\limits_{J=L+I_1+I_2}g_J T_L(E_{c.m.})
,\]
where $E_{c.m.}=\hbar^2k^2/2\mu,$ 
$g_J=(2J+1)/\{(2I_1+1)(2I_2+1)\},$
and $I_1$ and $I_2$ are the intrinsic spins of the incident particles. 

Examples of the NOC calculations for a number of lighter systems is given in
Fig. 23. The N/F values for the different systems are shown as a function of
$L_{gr}$.  Each of the curves shows a characteristic minimum at an $L_{gr}$
value that corresponds to an energy well above the corresponding Coulomb
barrier. The initial drop in N/F as a function of $L_{gr}$  is due to the
increasing difficulty as $L_{gr}$ increases of dissipating the angular momentum
brought into the  compound system solely through the evaporation of light
particles. The subsequent rise in N/F occurs when an increasing number of
direct channels (such as single and mutual inelastic excitation, nucleon and
$\alpha$-transfer and, finally, deep-inelastic orbiting and fusion-fission
processes) become accessible. These channels are activated at somewhat higher
energies because of their reaction $Q$-values. 

A strong correlation has been demonstrated between systems where N/F is small
and the occurrence of quasi-molecular resonances in light- and medium-light
heavy-ion reactions \cite{baa94}. For example, it is observed from Fig. 23 that
the  $^{12}$C+ $^{12}$C and  $^{12}$C+ $^{16}$O systems, where  prominent
resonant behavior \cite{baa94} has been observed, both show very low minimum
values of N/F $(N/F\approx 10^{-1}$). On the other hand, for systems such as
B+O, where the NOC values are much larger $(N/F>10^4)$,  there is no compelling
evidence for a strongly energy-damped reaction component other than what can be
accounted for through the fusion-fission mechanism \cite{aacf94}. 

Typical NOC calculations for a number of heavier systems with $36\leq
A_{CN}\leq 48$ are shown in Fig. 24.  These systems can be classified into two
groups, with the $^{24}$Mg+$^{24}$Mg reaction having an intermediate behavior.
The $^{24}$Mg+ $^{12}$C, $^{24}$Mg+$^{16}$O and $^{28}$Si+ $^{12}$C systems,
which are composed of  $\alpha$-particle-like nuclei, belong to the first group
and have NOC minima corresponding to $N/F\approx 10$.  For these systems,
strongly oscillatory angular distributions are observed in the backward-angle
elastic scattering cross sections \cite{baa94} and there is evidence of a
non-statistical orbiting mechanism \cite{s82,gdbh90,rk85}.   The second group
of reactions, with large values of N/F,  show little evidence of an excess
yield beyond that expected from the statistical fission mechanism. 

The $^{24}$Mg+$^{24}$Mg system appears to be much more surface-transparent at
large $L_{gr}$ in comparison to $^{23}$Na+ $^{24}$Mg, for instance, but never
achieves the low N/F values of the reactions populating the $^{40}$Ca compound
system. One distinguishing feature of the reaction is that the
``quasi-molecular resonance window", i.e., the  region of relatively low N/F
values, corresponds to quite high values of $L_{gr}$. Very narrow, high-spin
resonances have been observed in elastic and inelastic scattering measurements
for this system \cite{zkbsh83,qzkp90}  with strong correlation among the
various inelastic channels. These strong, heavy-ion resonance features seem to
coexist with ``normal", statistical fusion-fission yields from the $^{48}$Cr
compound nucleus,  as discussed in Ref. \cite{hsfp94}. The resonances, however,
 correspond to compound nucleus spin values \cite{qzkp90}  that approach the
N/F minimum, but are large compared to dominant spins expected to lead to
fusion-fission yields.  This circumstance of having a small number of open
channels for the grazing partial waves and, in addition,  having the angular
momentum characterizing these partial waves being greater than the dominant
angular momenta for the  statistical fission competition may contribute to the
strength of the observed resonance structure. 

\subsection{Dynamical model of intermediate mass fragment emission}

Fission in light systems can be considered as a dynamical process consisting of
the gradual shape change of the compound system with the formation of a neck
which subsequently narrows and breaks.  It is therefore tempting to include
dynamical effects into the statistical model calculations to have a better
understanding of the fission process and to provide some information on the
time scales which are involved during the reaction. 

The statistical model developed by Dhara et al. \cite{dbbk93}  is based on the
transition-state picture and involves solving the classical equations of motion
to follow the change of shape of the composite system. The fission dynamics are
therefore explicitly considered while evaluating the relevant physical
observables.  In the absence of any precise knowledge of the energy sharing
between the intrinsic excitation and collective degrees of freedom, it is
assumed that a random fraction of the initial compound-nucleus excitation
energy goes into collective degrees of freedom to generate the fission
dynamics. A standard proximity potential is used, with non-conservative
frictional forces introduced by a viscosity term whose magnitude has a temporal
dependence. The time evolution of this viscosity term has a strong impact on
the dynamics of the fission process.  The fission probability is calculated by
Monte Carlo simulation for a large number of trajectories. The spin transferred
to the binary fragments are computed in the sticking limit. The model still
uses phase space arguments to determine the fusion-fission cross sections,
however, in a manner that is quite similar to the transition-state model
approach \cite{s91,lgm75}. The two models differ primarily in their parametric
expressions describing the nuclear shapes and the saddle-point energies.
Calculations using the dynamical model have been shown to reproduce reasonably
well the experimental results obtained for the $^{31}$P compound system
\cite{bbb91,bbm95,3b96}. 

\subsection{Generalized liquid drop model}
\label{sec:ternary}

Within the macroscopic energy models, the fission saddle-point is found by
studying the potential energy surface of the compound system as a function of a
specific shape parameterization.  One such parameterization has been developed
by  Royer and Remaud \cite{rr84} involving a class of shapes that evolves from
a spherical mononucleus to two touching spherical fragments with a deeply
creviced neck. A nuclear proximity potential is used to account for the surface
interaction of the neck configuration. 

An interesting aspect of this parameterization is that it is easily generalized
to a three fragment configuration, allowing the study of a possible ternary
fission valley \cite{gr95}.  The possibility of a ternary fission path has been
explored through model calculations for the $^{48}$Cr compound system.  In this
study it is found that a prolate aligned $^{16}$O+ $^{16}$O+ $^{16}$O molecule
configuration results in a minimum of the potential energy as a function of
deformation for high spins.  Such behavior could result in an enhanced cross
section for ternary fission to the  three  $^{16}$O final channel. A shell
stabilized, three  $^{16}$O chain configuration of $^{48}$Cr has also been
suggested by unconstrained $\alpha$-cluster model calculations by Rae and
Merchant \cite{rm92}.  An experimental search for the three  $^{16}$O breakup
channel of $^{48}$Cr, as populated through the $^{24}$Mg+ $^{24}$Mg reaction,
has failed to find evidence for this behavior, however \cite{mmetc96}. 

\section{System-by-system Discussion}

The compound nuclei considered in this report range in mass from $19\leq
A_{CN}\leq 80$.  The low mass limit is set by the experimental difficulty of
unfolding the fully energy damped yields from the  more peripheral reaction
yields in even lighter systems.  The high end of the mass range corresponds to
the transition region where, for the partial waves that lead to fission in
heavy-ion reactions, the potential energy surface becomes relatively flat as a
function of mass asymmetry. For systems with mass lower than $A_{CN} \approx$
80, the potential energy surface favors fission of the compound system into
fragments of unequal mass.  For heavier systems, the symmetric mass fission of
the compound nucleus is favored in the absence of strong shell effects.  Tables
1 and 2 summarize the experiments that have been performed in lighter systems
where fully energy-damped yields have been studied. In the following
discussion, a somewhat arbitrary division is made between systems with $A_{CN}
< 32$,  $32 \leq A_{CN}\leq 44$,  $44<A_{CN}\leq 56$, and $56<A_{CN}\leq 80$.
The systems with $A_{CN} < 32$ are the most difficult ones in which to
experimentally establish the fission cross sections.  They also create the
greatest challenge to the fission-model calculations because of the difficulty
of calculating fission barriers for these very light systems and the increasing
influence of shell corrections on these barriers.  The mass range  $32 \leq
A_{CN} \leq 44$ encompasses the systems for which there is the greatest
evidence of a dinucleus orbiting mechanism that is distinctly different from
the expectations of the fission model calculations. In the mass region
$44<A_{CN}\leq 56$, there is good overall agreement of the expectations of 
fission transition state model with the observed fully energy damped yields,
although an additional heavy-ion resonance behavior is observed for several
systems in this mass range.  A shift to symmetric mass fission starts to become
evident towards the higher end of the mass region with $56<A_{CN}\leq 80$. 
\begin{table}
\caption{Systems studied with $A_{CN}\leq 40.$}
\begin{tabular}{|lccccccc|} \hline 
 $A_{CN}$& Reaction$^a$& $E_{lab}$& Detected&
Tech-&
C or& $NOC$ & Ref. \\
&& (MeV) &Fragments&nique&S$^b$&(min.)&\\ 
\hline
$^{19}$F&$^9$Be+$^{10}$B&10-40&$5\leq Z\leq 9$&
$\Delta$E-E&S&
$1.4\times 10^{-3}$ &\cite{faa93}\\
$^{20}$F&$^{9}$Be+$^{11}$B&10-40&$5\leq Z\leq 9$&
$\Delta$E-E&S&
$1.6\times 10^{-2}$&\cite{faa93}\\
$^{20}$Ne&$^{10}$B+$^{10}$B&15-50&$5\leq Z \leq 9$&
$\Delta$E-E&S&
$7\times 10^{-3}$&\cite{caaf91}\\
$^{21}$Ne&$^{10}$B+$^{11}$B&15-50&$5\leq Z\leq 9$&
$\Delta$E-E&S&
$2\times 10^{-3}$&\cite{caaf91}\\
$^{22}$Ne&$^{11}$B+$^{11}$B&15-50&$5\leq Z\leq 9$&
$\Delta$E-E&S
&$2\times 10^{-3}$&\cite{caaf91}\\
$^{26}$Al&$^{16}$O+$^{10}$B&22-64& $5\leq Z\leq 9$&
$\Delta$E-E&S&$2\times 10^1$&\cite{aacf94}\\
$^{27}$Al&$^{16}$O+$^{11}$B&22-64&$5\leq Z\leq 9$&
$\Delta$E-E&S
&$5\times 10^0$&\cite{aacf94}\\
&$^{17}$O+$^{10}$B&22-64&$5\leq Z\leq 9$&
$\Delta$E-E&S&$1\times10^2$&\cite{aacf94}\\
$^{28}$Al&$^{17}O+^{11}B$&22-64&$5\leq Z\leq 9$&
$\Delta$E-E&S&$9\times 10^{1}$&\cite{aacf94}\\
&$^{18}$O+$^{10}$B&22-63&$5\leq Z\leq 9$&$\Delta$E-E
&S&$5\times 10^2$&\cite{aacf94}\\
&$^{19}$F+$^9$Be&56&$5\leq Z\leq 9$&$\Delta$E-E&
S,C&$3\times 10^{-3}$&\cite{aacf94}\\
$^{29}$Al&$^{18}$O+$^{11}$B&22-64&$5\leq Z\leq 9$&
$\Delta$E-E&S,C&$3\times 10^1$&\cite{aacf94}\\
$^{31}$P&$^{12}$C+$^{19}$F&96&$3\leq Z\leq 11$&
$\Delta$E-E&S&$1\times 10^2$&\cite{3b96}\\
&$^7Li+^{nat}Mg$&47&$4\leq Z\leq 9$&$\Delta$E-E&S
&$7\times 10^3$&\cite{bbm95}\\
&$a+^{27}$Al&60&$4\leq Z\leq 8$ &$\Delta$E-E& S&
$<$100&\cite{bbb91}\\
$^{32}$S&$^{20}Ne+^{12}C$&50-80&Z=6&$\Delta$E-E
&S&$7\times 10^0$&\cite{sfgsd79}\\
$^{36}$Ar&$^{20}$Ne+$^{16}$O&70-160&$6\leq Z\leq 9$
&$\Delta$E-E&S&$1\times 10^1$&\cite{sdc83}\\
&$^{24}$Mg+$^{12}$C&90-126&$5\leq Z\leq 9$&
$\Delta$E-E&S&$1\times 10^1$&\cite{gdbh90}\\
$^{40}$Ca& $^{24}$Mg+$^{16}$O&75-115&$6\leq Z\leq 8$
&$\Delta$E-E&S&$9\times 10^0$&\cite{rk85,kbg92}\\
&$^{28}$Si+$^{12}$C&99-180&$5\leq Z\leq 8$&
$\Delta$E-E&S&$2\times 10^1$&\cite{s82,ssf84}\\
&$^{20}$Ne+$^{20}$Ne&70-160&$8\leq Z\leq 10$&
$\Delta$E-E&C&$7\times 10^1$ &\cite{sdc83}\\ \hline
\end{tabular}
$^a$ Projectile + Target \\
$^b$ C--Coincidence, S--Singles
\end{table}
\normalsize

\begin{table}
\caption{Systems studied with $A_{CN}>40$.}
\begin{tabular}{|lccccccc|} \hline 
 $A_{CN}$& Reaction$^a$& $E_{lab}$& Detected&
Tech-&
C or& $NOC$ & Ref. \\
&& (MeV) &Fragments&nique&S$^b$&(min.)&\\ 
\hline
$^{42}$Sc&$^{28}$Si+$^{14}$N
&100-170&$3\leq Z\leq 10$&$\Delta$E-E&
S&$8\times 10^3$&\cite{sss86}\\
$^{43}$Sc&$^{27}$Al+$^{16}$O&120-160
&$6\leq Z\leq 9$&$\Delta$E-E&S
&$7\times 10^2$&\cite{sfg80}\\
$^{44}$Ti&$^{24}$Mg+$^{20}$Ne&71-120&
$16\leq A\leq 32$&kin$^\dagger$
&C&$3\times 10^2$&\cite{bzb95}\\
&$^{28}$Si+$^{16}$O&61,79&
$6\leq Z\leq 8$&$\Delta$E-E&S&
$1\times 10^1$&\cite{oliv96}\\
&$^{32}$S+$^{12}$C&80-124&
$6\leq Z\leq 8$&$\Delta$E-E&S&$6\times
10^2$&\cite{ocd79}\\
$^{46}$V&$^6$Li+$^{40}$Ca&153&$9\leq Z\leq 11$&
$\Delta$E-E&S&$>$106&\cite{gmp84}\\
$^{47}$V&$^{23}$Na+$^{24}$Mg&89.1&$5\leq Z\leq 10$&
$\Delta$E-E&S&$4\times 10^4$&\cite{rsc91}\\
&$^{31}$P+$^{16}$O&135.6&$6\leq Z\leq 8$&
$\Delta$E-E&S&$1\times 10^4$&\cite{bdhf93}\\
&$^{35}$Cl+$^{12}$C&180-280&$5\leq Z\leq 12$&
$\Delta$E-E,&S,C
&$7\times 10^4$&\cite{bdhf92,bdhd89},\\
& & & &kin & & &\cite{alot96}\\
$^{48}$Cr&$^{24}$Mg+$^{24}$Mg&88.8&$6<A<24$&
kin&C&$5\times 10^2$&\cite{hsfp94}\\
&$^{28}$Si+$^{20}$Ne&87-128&$16\leq A\leq 32$
&kin$^\dagger$&C&$6\times 10^2$&\cite{bzb95}\\
&$^{20}$Ne+$^{28}$Si&78.0&$20\leq A\leq 24$&
kin&C&$6\times 10^2$&\cite{fsd96,cnrs94}\\
&$^{36}$Ar+$^{12}$C&187.7&$6\leq A\leq 24$&kin&
C&$3\times 10^3$&\cite{fsd96}\\
$^{49}$Cr&$^9$Be+$^{40}$Ca&141&$9\leq Z\leq 11$&
$\Delta$E-E&S&$>106$&\cite{gmp84}\\
$^{49}$V&$^{37}$Cl+ $^{12}$C&150,180&$5\leq Z\leq 11$
&$\Delta$E-E&S,C& $>106$&\cite{yno89}\\
$^{52}$Fe&$^{12}$C+$^{40}$Ca&74-186&
$9\leq Z\leq 11$&$\Delta$E-E&S&$8\times 10^4$&
\cite{gmp84}\\
$^{56}$Ni&$^{28}$Si+$^{28}$Si&85-150&A=28&
kin&C&$2\times 10^1$&\cite{betts81}\\
&$^{32}$S+$^{24}$Mg&121-142&$12\leq A\leq 28$&
tof,&S,C&$1\times 10^4$&\cite{setc89,shp94},\\
& & & &kin,$\gamma$ & & &\cite{sbjk90}\\
&$^{16}$O+ $^{40}$Ca&69-87&$20\leq A\leq 28$&
tof&S&$4\times 10^5$&\cite{sba86}\\
$^{59}$Cu&$^{35}$Cl+$^{24}$Mg&275&$5\leq Z\leq 12$&
$\Delta$E-E&S,C&$>10^6$&\cite{nbm96}\\
$^{60}$Ni&$^{16}$O+$^{44}$Ca&69-87&
$20\leq A\leq 30$&tof&S&$>10^6$&\cite{sba86}\\
$^{64}$Zn&$^{16}$O+$^{48}$Ti&118&$5\leq Z\leq 15$&
$\Delta$E-E&S,C&$>10^6$&\cite{yno89}\\
&$^{37}$Cl+$^{27}$Al&162-200&$5\leq Z\leq 15$&
$\Delta$E-E&S,C&$>10^6$&\cite{yno89}\\
$^{78}$Sr&$^{28}$Si+$^{50}$Cr&150&
$12\leq A\leq 58$&$\Delta$E-E,&S&$>10^6$&
\cite{esps89}\\
& & & & tof & & &\\
$^{80}$Zr&$^{40}$Ca+$^{40}$Ca&197,231&$7\leq A\leq 
62$&tof&S&$>10^6$&\cite{esps89}\\ \hline
\end{tabular}
$^a$ Projectile + Target \\
$^b$ C--Coincidence, S--Singles \\
$^\dagger$ Only low-lying excitation studied.
\end{table}

\subsection{$A_{CN} < 32$}

In describing heavy-ion reaction behavior in terms of statistical models, the
dominant component of the deformation-dependent binding energy of the compound
system is usually obtained through macroscopic energy calculations, that is,
extensions of the rotating liquid drop model.  Shell corrections to the
calculated  energies are usually handled in only a very approximate fashion.
These corrections can be difficult to obtain for the relevant shapes of the
compound system and, moreover, the statistical phase space relevant for the
fission process is generally required at an excitation energy of the compound
system where shell corrections may already be strongly attenuated. 

The lower mass limit for the validity of macroscopic description of light
nuclear  systems and statistical behavior of very light heavy-ion reactions has
been recently pursued by investigating reactions involving light s-d shell
nuclei.  One of the interesting features that can be probed by these studies is
the effect of the channel spin on light ion reaction mechanisms.  Among the
systems studied,  the choice of $^{10,11}$B nuclei was based on the fact  that
very high channel spins $\{ ^{10}$B($3^+$) and $^{11}$B($3/2^+$)$\}$ are
involved when compared to the values of the grazing angular momenta.  This has
two significant consequences:  1.) the phase space in the exit channel is
enlarged and 2.) the composite system  angular momentum is increased. 

There are, however, experimental difficulties encountered in the study of
lighter systems.  All of the model calculations require some estimate of the
distributions of incident orbital angular momenta contributing to the fully
energy damped binary yields.  This generally requires a measurement of the
fusion cross sections for the reactions---in the fission models, the damped
binary yields define the high end of the fusion partial wave distribution
whereas, for the orbiting models, the higher fusion partial waves are seen as
competing with the orbiting mechanism.  Identification of fusion reaction
yields involving very light heavy ion reactions is complicated, however, by the
difficulty of  identifying the corresponding evaporation residues.  Frequently,
a given element can be produced by  nucleon or massive transfer, sequential
decay of a compound nucleus or by a binary decay of the composite system. The
production of $^8$Be residues,  which subsequently decay by breaking apart into
two $\alpha$ particles,  may not be negligible. To unfold the different
processes it is necessary to look for  differences  in their kinematics. Here,
velocity spectra can be very useful. An example of the unfolding procedure is
shown in Fig. 25 for the $^9$Be+ $^{11}$B reaction at $E_{lab}$=37 MeV and
$\theta_{lab}=8^o$ \cite{faa93}. At lower recoil velocities the spectrum is
consistent with a statistical model calculation using the code LILITA
\cite{cbd84}, as shown by the solid curve. The additional component, shown by
the dashed curve, can be attributed to a more peripheral reaction mechanism. 

Strongly energy damped yields have been investigated in the
$^{10,11}$B+$^{10,11}$B reactions simultaneously with fusion-evaporation
residue yields \cite{caaf91}. For the $^{10,11}B+ ^{11}B$ channels, the
observed energy spectra and excitation functions of the evaporation residues
are consistent with the general fusion systematics in this mass range. However,
a significant inhibition of the fusion cross section is observed in the
$^{10}$B+ $^{10}$B entrance channel  (see Fig. 26). At least part of the
missing flux may be directed to strongly energy damped decay yields in the Z=5
channel, which are found to be strongly enhanced.  These yields demonstrate
binary characteristics according to their velocity distributions (which are
peaked at higher velocities than expected for evaporation residues) and
isotropic  angular distributions \cite{caaf91}. A similar enhancement in the
$^{11}$B channel is not  observed in either the $^{10,11}$B+ $^{11}$B or the
$^9$Be+ $^{10,11}$B reactions up to $E_{lab}/A$= 4 MeV. The mechanism by which
the very high entrance-channel spin of the $^{10}$B+ $^{10}$B system may play a
role in the anomalous behavior observed for this system is not clear.  Although
higher compound-nucleus spins would be expected to favor binary fission over
light particle emission, the behavior is also similar to that observed in
somewhat heavier systems (with zero channel spin) and attributed to a dinucleus
 ``orbiting" mechanism. 

The experimental situation can be somewhat simpler for reactions involving mass
asymmetric entrance channels.  Here the use of inverse kinematics can lead to
clear identification of the target-like products corresponding to processes
that are fully energy damped and emitted at large center-of-mass angles.  This
approach has been used in an investigation of the $^{16,17,18}$O+ $^{10,11}$B, 
$^{19}$F + $^9$Be reactions leading to the $^{26-28}$Al  compound nuclei
\cite{aacf94}. Contributions from binary reaction process were clearly
identified in the Li, Be, C, O and F channels.  However, only in the
target-like particle channels: Li, Be, B, and C , were isotropic distributions
of ${d\sigma / d\theta_{c.m.}}$ observed for all the reactions.  Alternatively,
the N, O and F products, associated with projectile-like particles, present
forward-peaked angular distributions with ``life-angles" in the range  $25^o <
\alpha < 50^o$  (see Fig. 6), suggesting that a more peripheral reaction
mechanism dominates the small-angle yields in these channels. The velocity
distributions of the emitted fragments suggest the occurrence of a  binary
process.  This is shown for the $^{18}$O+ $^{11}$B reaction at $E_{lab}$=63 MeV
in Fig. 27 from  ref. \cite{aacf94}.  Although the singles data offer
compelling evidence of  binary nature of the fully energy-damped yields, the
experimental confirmation of such a nature requires the detection of both
fragments in coincidence. Such coincidence measurements were performed for the
$^{18}$O+ $^{11}$B and $^{17}$O+ $^{11}$B reactions at $E_{lab}$=53 MeV, with
the observation that most of the yields correspond to $Z_1+Z_2=Z_{CN}$.  This
indicates that the effect of secondary light-particles emission following (or
preceding) scission is negligible for these reactions at the measured energy.
Coincidences  between heavy fragments and $Z<3$ particles were associated with
evaporation residues. 

The degree of inelasticity of the binary yields associated with distributions
of constant  ${d\sigma / d\theta_{c.m.}}$ is indicated by the total kinetic
energy $<TKE>$ values displayed in Fig. 28 for these yields. Compared to the
experimental results are lines showing the  expected values in the case of
totally relaxed processes in which the outgoing  particles carry essentially
the barrier energy. Although these results were calculated assuming spherical
fragments, similar results are found using the transition-state model for
fission where more realistic shapes are assumed for the compound  system
configuration \cite{aacf94}. 

Support for the picture of a statistically equilibrated compound nucleus is
found in the data obtained for the $^{17}$O + $^{11}$B, $^{18}$O + $^{10}$B and
$^{19}$F + $^9$Be reactions, each of which populates the $^{28}$Al compound
system.  Figure 29  presents excitation functions for intermediate mass 
fragment cross sections for the three systems. The  close agreement among the
exit yields for the various entrance channels shows that the Bohr Hypothesis
\cite{bw79} is satisfied. This hypothesis states that the exit channel
observables for compound-nucleus reactions should be independent of the
entrance channel except for such conserved quantities as angular momentum and
total energy. A similar conclusion is reached by comparing the ratio of yields
for different exit channels as a function of the excitation energy of the 
emitted fragments. Figure 30 presents the ratio $R=  \sigma_C / \sigma_B$ of
yield for carbon products $(\sigma_C)$ compared to the yield for boron
$(\sigma_B)$, for the three entrance channels populating the same compound
nucleus.   Again, entrance channel independence of the cross section ratios is
observed. Although it is not possible, in general, to form a compound nucleus
with the same excitation energy and angular momentum distribution using two
different entrance channels, the three reactions reaching $^{28}$Al  achieve
similar conditions because of  the small variation in the entrance channel mass
asymmetry. 

Predictions of the transition-state model for the fission charge distributions
are also compared to the experimental data in Fig. 29. In general, model
predictions are found to be satisfactory in reproducing the charge, mass and
bombarding energy dependence of the observed yields, thus further supporting
the idea that these yields have a fusion-fission origin. It can also be noted
that for these reactions, where the statistical fission description seems to
work well,  the normalized Number of Open Channels (N/F) for the composite
system decay are relatively high---some two to four orders of magnitude larger
than the ones obtained for the $ ^{12}$C+ $^{12}$C or  $^{12}$C+ $^{16}$O
resonant systems, as shown in Fig. 23. 

Bhattacharya {\it et al.} \cite{bbb91,bbm95,3b96} have looked for a possible
entrance channel behavior for the fully energy damped yields from the $^{31}$P
compound system as populated through the $^{12}$C+ $^{19}$F, $^7$Li+
$^{nat}$Mg, and $\alpha$+ $^{27}$Al reactions.  In this case there is a very
large difference in the mass asymmetries of the respective reactions.  Again,
in a comparison of the damped binary yields for the three systems, no prominent
entrance channel behavior is observed beyond what would be expected from the
very different partial cross-section distributions.    At the respective
bombarding energies, the normalized NOC for the grazing partial wave is large
for each of these systems. 

\subsection{$32 \leq A_{CN} \leq 44$}
\label{sec:a32to44}

The first observation of fully energy damped reaction yields in light systems
was reported by Shapira {\it et al.} \cite{sfgsd79} in an investigation of
$^{20}$Ne+ $^{12}$C inelastic scattering at backward angles. Large cross
sections were found when summing the yields over many unresolved mutual
excitations at  high excitation energies.  The resulting angular distributions
were found to show most of the characteristics, such as a ${d \sigma / d\Omega}
\propto{1 / \sin\theta_{c.m.}}$ angular dependence, of a long-lived, orbiting,
$^{20}$Ne+ $^{12}$C dinuclear complex. This initial measurement was followed by
a very detailed study of the same system \cite{sfg82} where  resonant-like
behavior was found in the excitation functions for several outgoing channels,
reminiscent to that observed for the well-known, resonant, symmetric  $^{16}$O+
$^{16}$O system. This  quasi-molecular resonance behavior extends to $Q$-value
regions where the total kinetic energy in the exit channel is consistent with
an orbiting-type mechanism. It was noted by Shapira {\it et al.} \cite{sfg82}
that the small number of open reaction channels for $^{20}$Ne+ $^{12}$C
scattering might be related to the observation of the resonance behavior. 

In a study of the $^{20}$Ne+ $^{16}$O reaction, Shapira {\it et al.}
\cite{sdc83} again found evidence of an ``orbiting-like" component in the large
angle $^{12}$C and  $^{16}$O yields. Subsequent to this,  the strongly energy
damped component in the $^{24}$Mg+ $^{12}$C reaction,  leading to the same
$^{36}$Ar composite system, was investigated by the Munich group through a
series of measurements \cite{gdbh90,dgh88}. In the study by Glaesner {\it et
al.} \cite{gdhk86}, cross section fluctuations for fully damped yields were
observed for the first time in such a heavy system, allowing for an
Ericson-type fluctuation analysis \cite{emk66}.  Excitation functions were
generated for different $Q$-value ranges, as shown for the Z=6 channel in Fig.
31, using the data of Ref. \cite{gdhk86}.  The observed structure is less
pronounced than that observed in discrete-level, low $Q$-value channels because
of the averaging over a large number of states. The coherence width obtained in
this measurement of approximately 300-400 keV  corresponds to a mean rotation
angle for an orbiting-like configuration of 180$^o$ to 360$^o$ .  In further
support of an orbiting-like picture, Konnerth {\it et al.} \cite{kdetc88} have
deduced spin alignments for the $^{24}$Mg+ $^{12}$C system by measuring the
out-of-plane $\gamma$-ray anisotropy. The  large positive values of the
observed spin alignments suggest the geometry of a sticking dinuclear complex
in a stretched configuration. 

Perhaps the most striking example of orbiting behavior in a light nuclear
system can be seen in the $^{28}$Si+$^{12}$C reaction yields.  This system has
been extensively studied over the past two decades for a number of reasons.  In
addition to showing a very  strong orbiting-like behavior
\cite{s82,ssf84,rsh91,rlv86} (see Fig. 22), it also demonstrates strong
resonant structures in its elastic and quasi-elastic yields
\cite{blb79,cfo78,ocd79}.  Fortunately, the reaction is favorable for
experimental study since intense $^{28}$Si beams are readily available at
tandem accelerator facilities and  $^{12}$C makes a very good and relatively
contaminant free target. 

The energy spectra for the most intense C, N and O fragment exit channels for
the $^{28}$Si+ $^{12}$C reaction have been measured at backward angles for a
large number of incident energies in the range $29.5\leq E_{c.m.}\leq 54 MeV$.
At lower bombarding energies the excitation spectra for the  $^{12}$C fragments
are dominated by single and mutual excitations of the  $^{12}$C and $^{28}$Si
fragments.  At higher bombarding energies, however, the dominant strength for
all three channels  shifts to higher excitation energies \cite{ssf84}.  For
these higher energy spectra the most probable $Q$-values are found to be
independent of detection angle. The corresponding, energy-integrated angular
distributions are found to follow a $1/\sin\theta_{c.m.}$ angular dependence
near 180$^o$\cite{sfg82}.  These  characteristics suggest a long lived,
orbiting-like mechanism where energy equilibration has been achieved. 

Many of the salient features of the orbiting yields, such as their inelasticity
and anisotropy, are indistinguishable, however,  from those expected for a 
compound nucleus fusion-fission mechanism.  Early on, the fission mechanism was
considered as a possible explanation for the energy damped $^{28}$Si+$^{12}$C
yields, but abandoned  because the observed cross sections were much greater
than model predictions based  on the standard rotating liquid drop model.  The
possibility of a significant  fission contribution was again suggested after 
it was shown that finite-nuclear range effects can lead to significantly
reduced fission barriers in lighter systems \cite{s91,skb87,skb88}.  However,
as shown in Fig. 22, even though the newer fission calculations may be able to
account for the observed cross sections in the nitrogen and oxygen channels,
the calculations significantly understate the  cross sections measured in the
carbon channel \cite{s91}. 

Good agreement with the observed fully damped cross sections, evaporation
residue cross sections, and average total kinetic energy values for the damped
yields has been obtained using the equilibrium model for fusion and orbiting
\cite{sah87,ass88}, as discussed in Sec. \ref{sec:eqorb}.  In particular, the
observed saturation trend of the TKE values as a function of the incident
energy is well described by this latter model, as  shown in Fig. 21. In the
equilibrium orbiting model, saturation occurs because a value of the orbital
angular momentum is reached, after dissipation, beyond which the formation of a
dinuclear complex is not allowed because of the  centrifugal repulsion. 

Additional studies support the idea that an orbiting-like mechanism dominates
that component of the carbon yield characterized by an angular dependence with
${d\sigma / d\Omega}\propto {1 / \sin\theta_{c.m.}}$. The entrance-channel
dependence of the process was demonstrated by Ray {\it et al.} \cite{rk85} by
forming the $^{40}$Ca nucleus with both the $^{24}$Mg+ $^{16}$O and $^{28}$Si+
$^{12}$C reactions at closely matched excitation energies and angular momenta.
Figure 32 shows the observed ratio of the oxygen to carbon cross section for
these reactions as a function of excitation energy.  If compound-nucleus
fission dominated the cross sections, this ratio would be expected to be very
similar for the two reactions. The observation of a strong entrance-channel
dependence of the ratios suggests a non-compound mechanism. It is interesting
to note that the entrance channel effect becomes smaller at larger excitation
energies.  This might suggest stronger fission competition for the more
strongly damped yields \cite{cnrs94}. Shiva Kumar {\it et al.} \cite{kbg92}
have  extended the entrance channel studies by comparing the carbon to oxygen
cross section ratios over an extended range of $^{40}$Ca excitation energies.
This study suggests an approach to an equilibrated compound nucleus ratio as
the beam energy increases. 

In another study that supports an orbiting-like explanation for the less damped
yields, the population of magnetic substates of  $^{12}$C$(2_1^+)$ and
$^{28}$Si states for the $^{12}$C( $^{28}$Si, $^{12}$C) $^{28}$Si reaction at
180$^o$ has been measured by Ray {\it et al.} \cite{rlv86} using $\gamma$-ray
angular correlation techniques. Qualitatively, the observed selective
population of the m=0 magnetic substate with respect to the beam axis agrees
well with a simple dinuclear sticking picture. A subsequent, more complete and
detailed experiment has been performed to measure density matrices of the 
$^{12}$C and $^{28}$Si excited states  \cite{rsh91}. The data are found
consistent with a dinuclear picture in which bending and wriggling motions are
the dominant spin carrying modes.  It is interesting to note that these
$\gamma$-ray angular correlation experiments using the $^{28}$Si+ $^{12}$C
orbiting reaction have been used to illuminate certain conceptual aspects of
quantum mechanics \cite{ray93}. 

In general, orbiting processes and resonant-like structures have been found to
coexist in collisions for which surface transparency is expected based on the
small number of open reaction channels \cite{abe95}. The $^{28}$Si+$^{12}$C and
$^{24}$Mg+$^{16}$O  reactions, for example, both show large ``orbiting" yields
as well as strong resonances at backward angles in their elastic, inelastic
\cite{blb79,kovar85}, and transfer exit channels \cite{kovar85}.  As seen in
Fig. 24, the normalized number of open reaction channels for both of these
reactions is very small.  The mass-symmetric $^{20}$Ne+ $^{20}$Ne reaction,
which leads to the same $^{40}$Ca composite system as $^{28}$Si+ $^{12}$C and
$^{24}$Mg+ $^{16}$O, has also been investigated \cite{sdc83} and found to
exhibit strong orbiting-type behavior. Again, this system is found to have a
small number of open reaction channels.  However, to our knowledge, this is the
only  ``orbiting" system to show almost structureless elastic excitation
functions \cite{bzm95}. 

The $^{28}$Si+$^{14}$N reaction is an important test case for exploring how
surface transparency, as reflected in the number of open channel calculations,
is related to the occurrence of orbiting and molecular resonance behavior.
Although this system is close in  mass to the $^{28}$Si+ $^{12}$C system, the
phase space available to reaction channels is very different for the two:
whereas the $^{28}$Si+ $^{12}$C reaction has only a few exit channels, the
$^{28}$Si+ $^{14}$N reaction is characterized by a large value of N/F (see Fig.
24). The observed orbiting-like cross sections for $^{28}$Si+ $^{14}$N reaction
\cite{ssa88,sss86} are significantly smaller than for the $^{28}$Si+ $^{12}$C
reaction.  The equilibrium model for orbiting  \cite{sah87}  has been found to
give a reasonably good description  \cite{ass88} of the ensemble of  the
experimental data \cite{ssa88,sss86} for this reaction, although fusion-fission
calculations performed at $E_{c.m.} = 40 MeV$ \cite{s91}  are found to
reproduce the overall experimental behavior with comparable success.  As a
consequence of, or at least coincident with, the increased reaction phase
space, the damped binary breakup yield behave more like ``normal" fission. 

Pronounced backward angle yields have also been observed in the $^{27}$Al+
$^{16}$O reaction at several $^{27}$Al bombarding energies by Shapira {\it et
al.} \cite{sfg80}. This is also a system where the number of open reaction
channels is large and the observed cross sections for the damped yields are
consistent with a fission interpretation. 

The symmetric and nearly symmetric mass binary decay of $^{44}$Ti has been
studied in inclusive and exclusive measurements of the $^{32}$S+ $^{12}$C 
reaction at 280 MeV bombarding energy \cite{pbb86}. Although substantial
post-scission  evaporation occurs at this high energy, it was possible to
extract the energy-damped reaction cross sections and $\langle TKE \rangle$
values for this system. The analysis of these data is one of the earlier
attempts to describe the damped binary yields in a light system in terms of a
fusion followed by fission picture. However, the experimental set-up was
designed for the detection of two comparable mass fragments, rather than the
unequal mass fragments expected to dominate the fission cross sections for a
system of fissility below the Businaro-Gallone point \cite{bg55}. Similar
studies \cite{gmp84} focusing on symmetric mass breakup yields have shown the
existence of fission-like yields for three other light systems in this mass
region: $^6$Li+ $^{40}$Ca, $^9$Be+ $^{40}$Ca and  $^{12}$C+ $^{40}$Ca, leading
to the $^{46}$V, $^{49}$Cr  and $^{52}$Fe compound nuclei, respectively. 

In a study of the $^{28}$Si+$^{16}$O reaction at $E_{c.m.}$=39.1 and 50.5 MeV,
Oliveira {\it et al.} \cite{oliv96} found fully energy damped yields which were
attributed to a deep inelastic scattering mechanism.  These data have also been
analyzed in terms of the transition state model and found consistent with a
fusion-fission mechanism \cite{sdt97}. 

In another study of the $^{44}$Ti compound system, Barrow {\it et al.}
\cite{bzb95} have found evidence of correlated resonance phenomena in
excitation functions of binary channels from the $^{24}$Mg+ $^{20}$Ne reaction.
 The data suggest that the observed resonances can be characterized by angular
momenta close to that of the grazing angular momentum in the entrance channel. 
This is taken to suggest a different origin for these structures than the very
pronounced resonances seen in elastic and inelastic scattering yields for the
$^{24}$Mg+ $^{24}$Mg reaction \cite{zkbsh83}, which seem to arise from spins
higher than the corresponding grazing angular momentum \cite{qzkp90}. The
$^{24}$Mg+ $^{20}$Ne study focuses on low-lying excitations and does not
address the question of whether this resonance system is also found to exhibit
enhanced orbiting-like yields in the more strongly energy-damped channels.  A
similar resonance behavior to that seen for the  $^{24}$Mg+$^{20}$Ne reaction
is also observed for the $^{28}$Si+ $^{20}$Ne reaction, as also studied by
Barrow {\it et al.} \cite{bzb95}. Again, the possibility of an orbiting-like
component has not been explored. 

\subsection{$44 < A_{CN} \leq  56$}
\label{sec:exp44to56}

For reactions populating compound nuclear masses in the range $44 < A_{CN} \leq
 56$ there is relatively little evidence for the pronounced orbiting-like
yields observed in some lighter systems.  In general, the experimental cross
sections for the strongly damped binary yields are in good agreement with
expectations based on fission-model calculations.  There is still, however,
evidence for heavy-ion resonance behavior  for some of the systems studied in
this mass range. 

To study the possible competition between the fission and orbiting mechanisms,
the population and decay of the $^{47}$V compound system has been extensively
studied through three different entrance channels: $^{35}$Cl+ $^{12}$C;
$^{31}$P+ $^{16}$O and $^{23}$Na+ $^{24}$Mg
\cite{bdhd89,bdhf92,bmetc95,alot96}. These systems cover a wide range of
entrance-channel mass asymmetries and therefore allow for a strong test of the
decoupling of the observed binary yields from the entrance channel, one of the
signatures that can be used to differentiate between the fission and orbiting
mechanism. 

The  binary decay properties of the $^{47}$V nucleus, produced in the
$^{35}$Cl+$^{12}$C reaction, have been investigated between 150 and 280 MeV by
means of a kinematics coincidence technique \cite{bdhd89,bdhf92,bmetc95}.  The
angular distributions of the lightest fragments are found to follow a
$1/\sin\theta_{c.m.}$ angular dependence.  This is shown in Fig. 33 for
fragments with $5\leq Z\leq 11$ from measurements at $E_{lab}=180$ MeV and 200
MeV. Distributions of $d\sigma/d\theta_{c.m.}$ are shown. The distributions are
found to be independent of the scattering angle for each exit channels
indicating that the lifetime of the dinuclear complex is comparable to or
longer than the rotational period. 

The binary nature of the reaction products has been clearly established with
the coincidence measurement.  Complete energy relaxation of the fragments with
$5\leq Z \leq 11$  is evident from the angle independence of their observed TKE
values, as shown in Fig. 33 \cite{bdhf92}. Although these results are obtained
using singles data,  equivalent results have been obtained in the coincidence
measurements. The averaged TKE values for all detected fragments vary little
with incident energy and the TKE value corresponding to a symmetric mass
breakup is close to the prediction of the revised Viola systematics
\cite{vkw85}, as shown in Fig. 8. 

To test the entrance-channel independence of the damped reaction yield for the
$^{47}$V system, back-angle  $^{12}$C and $^{16}$O yields have been measured in
the $^{31}$P+ $^{16}$O \cite{rsc91} and $^{35}$Cl+ $^{12}$C reactions
\cite{bdhd89}  at energies leading to  the same compound-nucleus excitation
energy of $E^*_{CN}= 59.0$ MeV and very comparable angular momenta. The
observed  $^{12}$C and  $^{16}$O cross sections are comparable for the two
systems and much smaller than those predicted by the equilibrium orbiting model
\cite{sah87,ass88}.  Also, the ratio of carbon to oxygen cross section, as
shown  in Fig. 34, has no significant entrance channel effect and is in general
agreement with the predictions of the transition-state model calculations
\cite{s91}.  A similar comparison has been done with the $^{35}$Cl+ $^{12}$C
and $^{23}$Na+ $^{24}$Mg \cite{bdhf93} reactions, populating the $^{47}$V
compound nucleus at $E^*_{CN}= 64.1$ MeV. Again, as shown in Fig. 34, the
observed behavior is in reasonable agreement with the expectations of the
transition-state description of fission \cite{bdhf93}. The elemental cross
sections for the $^{23}$Na+ $^{24}$Mg reaction are also in agreement with
expectations based on the fission picture, as shown in Fig. 19. 

One of the most striking phenomena observed in heavy-ion reaction studies is
the  pronounced resonance structures observed in elastic and inelastic
scattering of the $^{24}$Mg+$^{24}$Mg system \cite{zkbsh83}. Narrow structures
which are correlated in many channels and extending to high excitation energy
suggest that a very special configuration of the $^{48}$Cr compound system is
formed in this reaction.  By measuring the  $\gamma$-ray correlations with the
$^{24}$Mg fragments, it has been possible to deduce a resonance spin for at
least one of the observed structures that is greater than that of the grazing
angular momentum \cite{qzkp90,wzk87}. This is taken to suggest a very prolate
deformed configuration of the compound system leading to the resonance.  In
earlier measurements of the resonance behavior, peaks were also observed in
excitation-energy spectra for the $^{24}$Mg($^{24}$Mg, $^{24}$Mg) $^{24}$Mg
reaction up to energies where secondary $\alpha$-particle evaporation from the
fragments obscures any spectroscopic details.  This raises the question as to
whether the structure observed at higher excitation energies is somehow related
to the resonance phenomenon or, instead, is a feature of the fission decay of
the compound nucleus. As discussed in Sec. \ref{sec:ternary}, a possible
ternary fission mode for the $^{24}$Mg+$^{24}$Mg reaction, as suggested by
several model calculations, has also been sought for but not observed. 

To explore the relationship of the different reaction mechanisms influencing
the binary decay yields of the $^{48}$Cr compound system, the energy-damped
yields of the $^{36}$Ar(E$_{lab}$=187.7 MeV)+ $^{12}$C \cite{fsd96},
$^{20}$Ne(E$_{lab}$=78.0 MeV)+ $^{28}$Si \cite{fsd96}, and
$^{24}$Mg(E$_{lab}$=88.8 MeV)+ $^{24}$Mg \cite{hsfp94} reactions have been
studied. Each of these reactions populates the $^{48}$Cr compound nucleus at an
excitation energy of about 59.5 MeV.   In each case, the outgoing  fragments
were identified by measuring both fragments in a kinematic coincidence
arrangement.  The calculated mass distribution of the fully energy damped
yields with $6\leq A\leq 24 $ is in excellent agreement with the experimental
results for the $^{24}$Mg+ $^{24}$Mg reaction \cite{hsfp94}, with no
significant evidence for an excess yield in the entrance channel that might
suggest an additional, orbiting mechanism. The agreement is also reasonable
good for the $^{36}$Ar+ $^{12}$C entrance channel, although in this case the
experimental results show a somewhat greater mass asymmetry of the fission
fragments than predicted.    The use of the kinematic coincidence technique
allows for very good $Q$-value resolution in the final channels, as seen in
Fig. 35 where the excitation-energy spectra for the $^{24}$Mg+ $^{24}$Mg
channel is shown for each of the three entrance channels.  The experimental
results are indicated by the thick-line histograms.  It is evident from this
figure that the structure observed at higher excitation energy is correlated
for the three entrance channels, making it improbable that this structure is an
artifact of the resonance behavior. Rather, the structure seems to be  related
to the detailed level structure of the final nuclei. 

The thin-line histograms in Fig. 35 are obtained using the transition-state
model with the saddle-point method and applying the same procedure as discussed
for Fig. 16 to associate the flux at the saddle point with specific mutual
excitations of the fragments \cite{shp94,fsd96}. The fission picture can
account for most of the observed structures, although with some significant
discrepancies observed between the calculated and measured yields to low-lying
excitations populated through the $^{24}$Mg+ $^{24}$Mg and $^{20}$Ne+ $^{28}$Si
channels.  This must be expected since both of these reactions show evidence of
resonance behavior \cite{zkbsh83,bzb95} which can not be described by the 
transition-state picture. The energies corresponding to single and mutual
excitations of yrast levels and ground-state band members are shown at the
bottom of  Fig. 35. Although these excitations are found to contribute to the
observed structures, they do not appear to dominate the spectra.  Instead, the
calculations suggest that random groups of high channel spin excitations
account for the general appearance of the spectra. 

Instead of using the saddle-point calculation to predict, {\it a priori}, the
excitation energy spectra, Farrar {\it et al.} \cite{fsd96}  have also used
this method to explore the compound-nucleus spin distribution leading to the
observed fission yields.  In this analysis, it is found that the  average spin
obtained from the fitted distribution is comparable to that  obtained from the
{\it a priori} calculation for the $^{36}$Ar+ $^{12}$C and $^{20}$Ne+ $^{28}$Si
reactions, but is smaller than the systematics would suggest for the $^{24}$Mg+
$^{24}$Mg entrance channel. The difference in the average fitted and calculated
spin values is even greater when a ``resonance-subtracted" energy spectrum is
used for the $^{24}$Mg+$^{24}$Mg reaction. This suggests that there may be
direct competition between the heavy-ion resonance and compound-nucleus fission
mechanisms for near grazing partial waves of this entrance channel. 

The $^{56}$Ni compound system has also been explored through multiple entrance
channels, using the  $^{16}$O($E_{lab}$=69-87 MeV)+ $^{40}$Ca \cite{sba86},
$^{28}$Si($E_{lab}$=85-150 MeV)+ $^{28}$Si \cite{betts81}, and
$^{32}$S($E_{lab}$=121 MeV and 142 MeV) + $^{24}$Mg reactions \cite{setc89}.
Excitation functions of the elastic and inelastic excitations of the symmetric
$^{28}$Si+ $^{28}$Si channel are found to exhibit correlated resonance behavior
\cite{bdp81}, although somewhat more weakly than seen in the $^{24}$Mg+
$^{24}$Mg system. Each of the three entrance channels is found to result in
fully energy-damped yields that are consistent with a fusion-fission reaction
mechanism. 

Excitation-energy spectra for the $^{24}$Mg( $^{32}$S, $^{28}$Si) $^{28}$Si
reaction at $E_{lab}$=121 MeV and 142 MeV \cite{setc89} are shown in Fig. 17
and are found to be well reproduced by the transition-state model using the
extension to the saddle-point method \cite{shp94} to calculate these spectra. 
To further confirm the predicted population of final mutual excitation,
$\gamma$ rays were measured in coincidence with the $^{28}$Si fission fragments
\cite{shp94} in the excitation energy range $7.6\leq E_x\leq 16.7$ MeV. In
general, the observed and  predicted transition rates for specific states in
the $^{28}$Si fragments were found to be in good agreement. This is shown in
Figs. 36 and 37. In Fig. 36 it is seen that the model calculations well
reproduce the relative strength of the yrast $2^+\rightarrow 0^+$ and
$4^+\rightarrow 2^+$ transitions. Since higher spin states tend to feed the
$4^+$ level, this agreement suggests that the population of these higher spin
states is being reasonable well described. Figure 37 shows the calculated and
measured $\gamma$-ray spectra on an expanded scale. Again, reasonably good
agreement is found between the predicted and observed transition strengths.  A
quantitative comparison, however, reveals evidence of greater population of
members of the $K^\pi=0^+_3$ band in $^{28}$Si (taken to include levels at 6691
keV, 7381 keV, 9165 keV, and 11509 keV) than predicted.  This band is believed
to have a strongly prolate deformed nature \cite{gbs81} and an enhanced
population might result from the relatively  deformed shapes expected for the
nascent fission fragments at the saddle point. It should be noted, however,
that there is some debate as to whether  the 11509 keV $6^+$ level should be
associated with this band \cite{cshs82}. 

The mass distributions of the fission-like yields for the $^{32}$S+ $^{24}$Mg
reaction at $E_{lab}$=121 and 142 MeV \cite{setc89}, obtained by fitting
experimental angular distributions assuming a ${d\sigma / d\Omega} \propto {1 /
\sin\theta_{c.m.}}$ angular dependence,  are shown in Fig. 38 by the open
histograms.  The corresponding predicted mass distributions based on the
transition-state model calculations, shown by the solid histograms, are found
to be in excellent agreement with the data.  One of the results of the model
calculations which has yet to be confirmed in any of the systems studied is
that significant fission yield is expected in the $^8$Be channel.  By bridging
the light-particle evaporation and fission mass ranges, the predicted strength
in the $^8$Be channel highlights the idea, as proposed by  Moretto
\cite{lgm75}, that light-particle evaporation and fission yields have a common
origin and should be viewed in terms of their respective decay barriers. 

In an experiment exploring the energy and spin sharing between fission
fragments in the $^{24}$Mg( $^{32}$S, $^{12}$C) $^{44}$Ti reaction at
$E_{lab}$=140  MeV, $\gamma$-ray spectra were obtained in coincidence with the 
 $^{12}$C fragments \cite{sbjk90}.  The results also indicate a statistical
decay process consistent with the predictions of the transition-state model. 
Moreover, no evidence was found for the spin alignment of the  $^{12}$C
fragments, contrary to what might be expected for a deep-inelastic scattering
origin of the fully energy-damped yields. 

\subsection{$56 < A \leq 80$}

The reaction products from the $^{35}$Cl+ $^{24}$Mg system have been
investigated at a bombarding energy $E = 8$ MeV/nucleon with both inclusive
\cite{cbd95} and exclusive measurements \cite{nbm96}. The inclusive data
provide information on the properties of both the evaporation residues and the
binary-decay fragments. The binary process yields are, for  instance,
successfully described by statistical models based on either the saddle point
picture \cite{s91} or the scission point picture \cite{mbnm97}. The similar
good agreement with theory that is found for the energy spectra, the angular
distributions, and the $\langle TKE \rangle$ values makes the hypothesis that
fully energy-damped fragments result from  a fusion-fission process quite
reasonable for the $^{59}$Cu compound system, in accordance with findings for
other, equivalent systems as shown in this report. 

No evidence is seen in the coincidence data for the occurrence of three-body
processes in the $^{35}$Cl+ $^{24}$Mg reaction.  This result can be contrasted
to the situation reported for somewhat heavier mass systems, where significant
three-body breakup yields are evident for the $^{32}$S+ $^{45}$Sc \cite{bmv93}
and $^{32}$S+ $^{59}$Co \cite{vml88} reactions, both measured with
$E_{lab}$($^{32}$S)=180 MeV (5.6 MeV/u). The nuclear-charge deficits from the
compound-nucleus charge found in the $^{35}$Cl+ $^{24}$Mg exclusive
measurement, however,  can be fully accounted for by the sequential evaporation
of light charge particles (LCP), in agreement with the systematics established
for a large number of reactions studied at bombarding energies below 15
MeV/nucleon \cite{bsdt96}. The question of whether a small part of the binary
reactions come from ternary processes is still an open question and difficult
to answer.  In general the measured charge-deficit values and other
experimental observables (such as cross sections, energy- and
angular-distributions or mean TKE values) are very well described by a complete
Extended Hauser-Feshbach statistical-model calculation \cite{mbnm97} which
takes into account the post-scission LCP and neutron evaporation. 

Although the fission picture is seen to work well in the $^{35}$Cl+ $^{24}$Mg
reaction,  a very different conclusion is reached by Yokota {\it et al.}
\cite{yno89} in a study of two systems ($^{37}$Cl+ $^{27}$Al and $ ^{16}$O+
$^{48}$Ti) leading to the somewhat heavier $^{64}$Zn compound system.  These
reactions populate  $^{64}$Zn at comparable excitation energies and spins.  The
components of the reaction yields corresponding to a ${d\sigma /
d\Omega}\propto {1 / \sin\theta_{c.m.}}$ angular dependence result in very
different $Z$ distributions for the two reactions, contrary to the expectations
of the statistical decay of a compound nucleus.  It is possible that these
systems are again displaying the strong orbiting signature that has been found
in several lighter systems.  If so,  these results could provide an interesting
challenge for the number of open channels calculations performed for systems of
mass $A_{CN} > 60$. 

The transition region where it appears that the asymmetric-mass fission
observed in lighter systems may change into the symmetric fission behavior
characteristic of heavier systems appears to occur around mass $A\approx 80$.  
Evans {\it et al}.  have studied the $^{40}$Ca+ $^{40}$Ca reaction at
$E_{c.m.}$=197 MeV and 231 MeV \cite{esp91} and the $^{28}$Si+ $^{50}$Cr
reaction at $E_{lab}$($^{28}$Sr)=150 MeV \cite{esps89}. The resulting mass
distributions for the lower energy $^{40}$Ca+ $^{40}$Ca reaction and the
$^{28}$Si+  $^{50}$Cr reaction are shown in Fig. 39 by the open circles.  The
mass distributions for the two reactions are found to be quite different, even
though the fissility of the two systems is quite similar.  This behavior was
initially thought to indicate a fast-fission mechanism accounting for the
asymmetry dependence of the $^{28}$Si+ $^{50}$Cr yields \cite{esps89}. However,
in exploring the possibility of fission competition in these systems, it has
been found that for the spin values near the critical angular momentum for
fusion, where most of the fission yields is expected to originate, the mass
asymmetry dependent fission barriers are actually quite different for the two
systems. The $^{80}$Zr compound system, populated through the
$^{40}$Ca+$^{40}$Ca reaction, is found to have either a very flat distribution
of barrier energies as a function of mass asymmetry or slightly lower barriers
for  the symmetric mass configuration. For the $^{78}$Sr compound system,
populated through the $^{28}$Si+ $^{50}$Cr reaction, however, the distribution
favors asymmetric mass breakup. The bold-line histograms in Fig. 39 show the
predicted primary fission mass distributions using the transition-state model
and the thin-line histogram show the corresponding mass distributions after
secondary light-particle  emission from the fission fragments is taken into
account.  For the two systems, the overall trend of the data seems to be
reproduced by the calculations, indicating that these two systems may straddle
the Businaro-Gallone \cite{bg57}  transition from asymmetric- to symmetric-mass
fission. 

\section{Open Problems}

\subsection{Time scale}

Although the process of binary decay from equilibrated compound nuclei has been
clearly identified experimentally and successfully described in terms of
phase-space models, more thought is needed on the dynamics through which
systems evolve from their entrance channels to  deformed and statistically
equilibrated compound nuclei that subsequently undergo scission. The occurrence
of orbiting and resonance behavior in some systems indicates the superposition
of compound nucleus decay products (properly described by the transition state
model) and faster direct processes (interpreted on the basis of polarization
potentials, entrance channel resonances or DIC orbiting processes). Therefore,
the investigation of the dynamics involved in heavy ion collisions leading to
totally energy-damped binary exit channels may lead to a better understanding
of the competition among the different reaction mechanisms. Within this
scenario two questions arise: a) the time scale of the processes and b) the
shape evolution of the system. 

It has been suggested by Thoennessen {\it et al.} \cite{tbb93} that systems
with different entrance channel mass asymmetries may evolve towards their
compound nucleus configurations with different time scales such that the most
asymmetric one reaches full equilibrium faster. This finding suggests that the
mass asymmetry may affect not only the angular momentum distribution but also
the competition with faster direct processes. In the case of heavier systems,
$\gamma$ rays from the decay of the giant dipole resonance (GDR) built on
highly excited compound-nucleus states has been shown to be sensitive (through
the shape of the energy spectrum) to the time scale of the process as well as
to the deformation of the nuclear system. As we go to lighter systems, where
the energy of the GDR is quite high, severe experimental problems are expected
with such GDR measurements. However, these measurements should be extended to
systems that are as light as possible. 

Interferometry measurements based on the detection of emitted pairs of light
charged particles has been employed as a technique to probe time scales and
nuclear dimensions \cite{koon77,dgx89}. For heavier systems, these studies have
generally involved the use of small angle correlations of protons emitted from
a hot source in the determination of emission time scales and sources radii.
However, if we go to lighter and equilibrated systems, this technique can be
borrowed to obtain estimates of reaction time scales. If a hot composite
systems decays into a binary channel and a proton or alpha particle evaporates
from one of the fragments, the proximity of the other fragment may  distort the
kinematics correlation  because of the strong Coulomb repulsion. Such a
distortion will depend on the time scale for fission and the sequential
secondary evaporation. 

In cases where there is good $Q$-value resolution, the experimental $Q$-value
spectra present a structured behavior even at high excitation energies.  This
is seen, for example, in Fig. 35 for the three entrance channels populating the
$^{48}$Cr compound nucleus. These structures can be associated with selectively
populated clusters of high-spin mutual excitations.  This situation allows for
a fluctuation analysis (see, for example, ref. \cite{gdhk86} and Sec.
\ref{sec:a32to44}) to obtain the coherence width of the intermediate system
and, hence, its lifetime (with $\tau=\hbar / \Gamma$). It is important to
assure that the number of levels included in the energy bin is very low,
requiring experiments using very thin targets and beams of good energy and
spatial resolution. 

\subsection{ Relationship to heavy-ion resonance behavior-superdeformed
minima?} 

Several of the results presented in this report suggest that a coherent
framework may exist which connects the topics of heavy-ion molecular resonances
\cite{eb85,gps95}, superdeformation effects as observed in medium mass
$\gamma$-ray studies (see the most recent reports quoted in \cite{jk91}), and
fission shape isomerism in the actinides \cite{vh73}. 

The shape of the ``normal" saddle-point configuration in light systems is very
similar to two, touching, prolate-deformed spheroids in a neck-to-neck
configuration. The shape of the system found in a conjectured, secondary well
in the potential energy surface is likely to be similar, although involving
greater deformation of the nascent fragments. 

In calculations of the shape-dependent potential-energy surfaces at high
angular momenta (16-40$\hbar$) for the $^{48}$Cr  nucleus \cite{aberg90}, a
strong superdeformed configuration is predicted that corresponds to an aligned
arrangement of two touching and highly deformed $^{24}$Mg nuclei. This
superdeformed configuration is a candidate to become yrast at around spin
34$\hbar$,  in the high excitation energy region which corresponds to where the
quasi-molecular resonances have been observed. 

Indeed,  from spin alignment measurements \cite{qzkp90}  of two strong
resonances in the $^{24}$Mg+ $^{24}$Mg scattering reaction \cite{zkbsh83},  the
deduced spin assignments were found to be comparable to or a few units larger
than expected for grazing collisions, leading to the same conclusion that the
resonance configurations correspond to a shape of two prolate $^{24}$Mg nuclei
placed pole to pole. 

This observation has been further supported by theoretical calculations of a
molecular model \cite{ua93}  which indicate a dinuclear nature of the observed
resonances and suggest the presence of such a stabilized configuration in
$^{48}$Cr at high spins. The conjectured isomeric configuration constituting
this aligned, pole-to-pole arrangement of two $^{24}$Mg clusters has a large
probability for breakup into two $^{24}$Mg fragments---a situation which is
similar to that expected for the symmetric-fission saddle point, suggesting a
relationship between the fission mechanism and that responsible for the
resonance behavior. 

The relative strength of the various statistical and non-statistical processes
observed in the binary yields of light systems is found to be related to the
number of available open channels for the near-grazing partial waves
\cite{baa94,bmetc95}. The resonant, non-statistical mode of the $^{24}$Mg+
$^{24}$Mg reaction leading to the $^{48}$Cr compound system emphasizes the
dominance of partial waves near or slightly above the grazing angular momentum
value \cite{bdp81,zkbsh83}. The fission mechanism has also been found to play a
significant role at these spins \cite{hsfp94}, however, indicating that the
nuclear configuration leading to the resonance behavior is only slightly more
extended than that expected for the nuclear saddle point. 

The coexistence of fission and a separate reaction mechanism corresponding to
heavy-ion resonance behavior has been analyzed in detail \cite{hsfp94}  for the
binary breakup of the $^{48}$Cr compound system populated with the $^{24}$Mg+
$^{24}$Mg, $^{20}$Ne+ $^{28}$Si and $^{36}$Ar+$^{12}$C reactions (see Sec.
5.3). The conclusion drawn from inspection of the energy spectra shown in Fig.
35 was that a significant fraction of the yield observed in the
$^{24}$Mg+$^{24}$Mg exit channel arises from a statistical fission mechanism,
with the resonance mechanism primarily influencing the lower excitation energy
region of the spectra due to the more symmetric entrance channels. 

The influence of the fragment structure and the relationship between the
fission mechanism and that responsible for the resonance behavior needs to be
investigated further with detailed particle-$\gamma$  coincidence measurements
of the $^{24}$Mg+ $^{24}$Mg system, including excitation functions measurements
in the vicinity of the well-known resonance energies. 

\section{Conclusions}

In this review we have summarized the results and conclusions of many
investigators who have studied the fully energy-damped, binary yields arising
from reactions involving lighter nuclear systems with $A_{CN} \leq 80$. The
experimental and theoretical techniques used in these investigations have been
presented  and illustrated with experimental results. The general systematics
that have been developed for these yields have been reviewed. 

In general, the data lend support to the newer macroscopic energy calculations
based on the finite-range, rotating liquid drop model.  Fission like yields
have been observed in all of the systems studied. Moreover, the experimental
systematics support the expectation, based on model calculations, that fission
should favor a mass asymmetric breakup of the compound nucleus in these light
systems.  Further support for the fission picture comes for the measured total
kinetic energy values of the fragments which are found to reflect the expected
deformation of the compound nucleus at the point of scission. 

It is shown that various model calculations that share the premise that the
final mass and energy distributions can be described by phase space constraints
all lead to comparable predictions of the damped binary yields.  These include
the transition-state models based on counting the available states at the
saddle- or scission-points, the equilibrium orbiting model, and the dynamical
breakup model.  The strength of the transition-state model using the
saddle-point method is found in its ability to describe the general fission
behavior over the entire region of mass covered by this report in a relatively
``parameter-free" manner. This general success is believed to be related to the
ability of the finite-range rotating liquid drop model to correctly calculate
the shape and energy of the saddle-point barrier.  The similarity of the
saddle- and scission-point configuration in these light systems, however,
allows for very similar behavior being predicted when the scission point is
used as the ``transition state" or when equilibrium orbiting is considered. 

The general success of the statistical model calculations allows us to now
establish a reference for what is the ``expected" behavior for the damped
binary yields and to search for deviations from this behavior. It has become
clear that in some systems there is an additional ``orbiting" component that is
of much larger cross section than can be accounted for by the fission
calculations.  Systems where this additional component is present also tend to
manifest resonance-like behavior in excitation functions of their elastic,
inelastic, and transfer channels.  The occurrence  ``orbiting/resonance"
behavior is found to be strongly correlated with the number of open reaction
channels which, in turn, is believed to be associated with the degree of
absorption in the grazing partial waves.  The precise mechanism(s) involved in
the orbiting and resonance behavior is still unknown. 

The model calculations also make definite predictions of compound nucleus
lifetimes and the shapes corresponding to the fission saddle-point. 
Measurements aimed at confirming these predictions are likely to require triple
coincidences of three outgoing particles---the two resulting heavy fragments
from the reaction and either the $\gamma$ ray  or light particle emitted from
one of the two primary fragments.  Such measurements are still in their
infancy. 

\leftline{\bf Acknowledgements :}

This work was supported by the Conselho Nacional de
Desemvolvimento Cientifico e Tecnologico (CNPq), Brazil, 
the U.S. Department of
Energy, Nuclear Physics Division, under Contract No. DE-
FG03-96ER40981,  the
Centre National de la Recherche Scientifique of France 
within the CNRS/CNPq
Collaboration program 910100/94-0, and the U.S. National 
Science Foundation
under the U.S.-France and U.S.-Brazil International 
Programs offices.

\begin{figure}
\caption{Comparison of the fusion-fission (left branch) and dinucleus orbiting
(right branch) mechanisms.} 
\label{fig1}
\end{figure}

\begin{figure}
\caption{Elastic scattering distribution for the $^{16}$O+$^{28}$Si reaction
showing the ``anomalous" large-angle scattering behavior. The dashed curve is
the prediction of an optical model calculation. The insert compares a
$|P_\ell(cos\theta)|^2$ angular dependence to the large angle data.  (Figure
adopted from ref.[16]} 
\label{fig2}
\end{figure}

\begin{figure}
\caption{ Schematic  description of nuclear orbiting in heavy-ion reactions
[21]. } 
\label{fig3}
\end{figure}

\begin{figure}
\caption{ Fission-like  cross sections for the $^{35}Cl+^{12}C$ reaction at
$E_{lab}= 180$ MeV. This figure is reproduced from ref.[30]. } 
\label{fig4}
\end{figure}

\begin{figure}
\caption{ Angular distribution for the $^{20}Ne$ channel of the
$^{35}Cl+^{12}C$ reaction at $E_{lab} = 180$ MeV  (data taken from ref.[30]
together with a $1/\sin \theta$ angular dependence (dashed curve) and the
transition-state distribution (solid curve). The curves have been scaled to the
data. } 
\label{fig5}
\end{figure}

\begin{figure}
\caption{ Angular distribution anisotropy $R(\theta)$  as a function of the
``life" angle $\alpha$  for $\theta = 10^o$ and 170$^o$. The dashed line is the
corresponding anisotropy for a $1/\sin \theta$ dependent angular distribution.} 
\label{fig6}
\end{figure}

\begin{figure}
\caption{ Angular distributions of binary decay channels for the
$^{16}$O+$^{11}$B reaction.  An evaporation-residue component has been
subtracted to obtain these yields (see ref.[31]). The curves
represent fits using Eqn.(2.2) with the indicated ``life angles".}
\label{fig7}
\end{figure}

\begin{figure}
\caption{ Most probable TKE release in fission as a function of the parameter
$Z^2/A^{1/3}$ of the fissioning nucleus. Open triangles have been taken from
previous existing compilations [40,41].  Experimental solid points have been
recently compiled in ref.[42]. The dashed line is the result of the  Viola
systematics [41]. The solid line shows the revised systematics of Tavares and
Terranova [42].} 
\label{fig8}
\end{figure}

\begin{figure}
\caption{ Definition of variables used in determining fragment masses and
reaction $Q$-value for a  binary reaction using the kinematic coincidence
technique.} 
\label{fig9}
\end{figure}

\begin{figure}
\caption{ Experimental arrangement for studying binary yields from the
$^{24}$Mg+ $^{24}$Mg reaction [46]. }
\label{fig10}
\end{figure}

\begin{figure}
\caption{ Schematic of the reaction kinematics for a binary breakup process. } 
\label{fig11}
\end{figure}

\begin{figure}
\caption{ Comparison of the saddle-point energies deduced from the double
spheroid approximation (curves) and the corresponding full macroscopic energy
calculations (symbols),  at approximately 0.6 and 0.8 times the spin at which
the respective fission barriers vanish [39].} 
\label{fig12}
\end{figure}

\begin{figure}
\caption{ Saddle point shapes for the$^{40}$Ca, $^{56}$Ni, and $^{90}$Zr at the
indicated mass asymmetries and spins (solid curves).   The corresponding shapes
found for the double spheroid approximation discussed in the text are shown by
the dashed curves [39]. } 
\label{fig13}
\end{figure}

\begin{figure}
\caption{ Schematic diagram of the energy balance for a fusion-fission
reaction. } 
\label{fig14}
\end{figure}

\begin{figure}
\caption{ Saddle-point energies for the $^{56}$Ni compound nucleus as a
function of spin and mass asymmetry.  The mass-asymmetry coordinate is given by
the final fragment mass assuming fission occurs.  Typical saddle-point shapes
are also indicated. } 
\label{fig15}
\end{figure}

\begin{figure}
\caption{ Partial fusion cross sections as a function of compound nucleus spin
for the $^{32}$S+ $^{24}$Mg reaction at $E_{lab}$($^{32}$S)=121 MeV. The shaded
region indicates the cross section leading to fission. } 
\label{fig16}
\end{figure}

\begin{figure}
\caption{ $^{28}$Si+$^{28}$Si excitation-energy spectra for the $^{32}$S+
$^{24}$Mg reaction at (a) $E_{c.m.} =51.0$ MeV and (b) $E_{c.m.}=54.5$ MeV. 
The bold-line histograms are the experimental spectra.  The lighter curves are
the results of model calculations discussed in the text for fission decay to
particle-bound states (solid curve) and for all decays (dotted curve). This
figure is reproduced from ref.[49]. } 
\label{fig17}
\end{figure}

\begin{figure}
\caption{ Average calculated valued for the compound-nucleus spin $\ langle
J_{CN} \rangle$, exit-channel orbital angular momentum $\langle \ell_{out}
\rangle$, and channel spin $\langle s \rangle$ for the $^{28}Si+ ^{28}Si$
fission channel of the  $^{32}S+ ^{24}Mg$ reaction at $E_{c.m.}=51.0$ MeV. This
figure is reproduced from ref.[49]. } 
\label{fig18}
\end{figure}

\begin{figure}
\caption{ Experimental evaporation-residue and fission cross sections for the
$^{35}$Cl+ $^{12}$C reaction at $E_{lab}=200$ MeV and the $^{23}$Na+ $^{24}$Mg
reaction at $E_{lab}=89$ MeV (data taken from ref.[60]. The solid-line
histograms show the calculated cross sections based on the transition-state
model.  The dotted-line histograms are the comparable calculations using the
Extended Hauser-Feshbach model. The calculated fission cross sections have been
corrected for the influence of secondary light-fragment emission from the
fragments. } 
\label{fig19}
\end{figure}

\begin{figure}
\caption{ Experimental values of $\langle TKE\rangle$ for the $^{35}$Cl+
$^{12}$C reaction (squares, from ref.[30] at $E_{lab}=200$ MeV and the
$^{23}$Na+ $^{24}$Mg reaction (circles, from ref.[60]) at $E_{lab}=89$ MeV. 
The solid- and dashed-line curves show the expected values based on the 
transition-state model for the two systems,  respectively.  The dotted-line
curve show the expected values based on the Extended Hauser-Feshbach model for
the $^{35}$Cl+ $^{12}$C reaction. } 
\label{fig20}
\end{figure}

\begin{figure}
\caption{ Most probable experimental TKE values [17,61] of the orbiting
products from the $^{28}$Si+$^{12}$C reaction as compared to the equilibrium
model for orbiting [63] (dashed curve) and the transition-state model (solid
curves). } 
\label{fig21}
\end{figure}

\begin{figure}
\caption{ Experimental orbiting cross sections measured in the $^{28}$Si+
$^{12}$C reaction [17,61] as compared to the equilibrium model for orbiting
[63] (dotted curves) and the transition-state model (solid curves). } 
\label{fig22}
\end{figure}

\begin{figure}
\caption{ Calculated NOC values as a function of the grazing angular momenta
for selected light systems with $A_{CN}< 30$ [31]. } 
\label{fig23}
\end{figure}

\begin{figure}
\caption{ Calculated NOC values as a function of the grazing angular momenta
for selected light systems with $36 \leq A_{CN}\leq 48$ [18]. } 
\label{fig24}
\end{figure}

\begin{figure}
\caption{ Experimental velocity  spectrum for the boron elements detected from
the $^9$Be+$^{11}$B reaction with $E_{lab}=37$ MeV and $\theta_{lab}=8^o$.  The
curves are discussed in the text.  The figure is from ref.[80]. } 
\label{fig25}
\end{figure}

\begin{figure}
\caption{ Fusion excitation functions for the indicated  systems.  The dashed
curves show Glas and Mosel parameterization of the data.  The solid,  straight
lines are predictions of the total reaction cross sections based on optical
model calculations.  The solid curves  are based on proximity potential fits. 
This figure is reproduced from ref.[82]. } 
\label{fig26}
\end{figure}

\begin{figure}
\caption{ Velocity plot for the ``non-evaporation-residue" components of the
$^{18}$O+ $^{11}$B reaction at 63 MeV.  The circles, centered at the
center-of-mass velocity, describe the loci for products with constant $Q$
values.  This figure is reproduced from ref.[31]. } 
\label{fig27}
\end{figure}

\begin{figure}
\caption{ Excitation functions for the  average total kinetic energies of the
indicated systems and charge channels. The lines are based on barrier
calculations assuming spherical fragments.  This figure is reproduced from
ref.[31]. } 
\label{fig28}
\end{figure}

\begin{figure}
\caption{ Fission cross section excitation functions for the indicated systems
and charge channels. The three systems populate the common $^{28}$Al compound
system.  The dashed curves show the predicted cross sections based on the
transition-state model of fission.  This figure is reproduced from ref.[31]. } 
\label{fig29}
\end{figure}

\begin{figure}
\caption{ Ratio of carbon to boron yields as a function of excitation energy
for three different entrance channels populating the common $^{28}$Al compound
system.  This figure is reproduced from ref.[31]. } 
\label{fig30}
\end{figure}

\begin{figure}
\caption{ Excitation functions for the Z=6 channel of the $^{24}$Mg+ $^{12}$C
orbiting component at 5 different 1 MeV large $Q$-value windows around a) -12
MeV, b) -13 MeV, c) -14 MeV and d) -15 MeV [87].  } 
\label{fig31}
\end{figure}

\begin{figure}
\caption{ Ratio of the oxygen to carbon cross sections as a function of
excitation energy for the $^{28}$Si($E_{lab}$=115 MeV)+$^{12}$C and
$^{24}$Mg($E_{lab}$=79.5)+ $^{16}$O reactions [26]. The solid line shows
the expected ratio for both systems based on the transition-state model [96]. }
\label{fig32}
\end{figure}

\begin{figure}
\caption{ Angular dependence of the average TKE values measured in the
$^{35}$Cl+ $^{12}$C at $E_{lab}$ = 180 MeV and 200 MeV. This figure is
reproduced from ref.[30]. } 
\label{fig33}
\end{figure}

\begin{figure}
\caption{ Ratio of oxygen to carbon yields for entrance channels populating the
common $^{47}$V  compound nucleus.  The top panel shows the ratios for two
reactions populating the compound nucleus a an excitation energy of
$E^*_{CN}=59.0$ MeV. The reactions shown in the bottom panel populate the
compound nucleus at of $E^*_{CN}$=64.1 MeV. The dotted lines indicate the
predicted ratios based on the transition-state model calculations. The figure
is from ref.[60]. }
\label{fig34}
\end{figure}

\begin{figure}
\caption{ The $^{24}$Mg+ $^{24}$Mg energy spectra for the $^{36}$Ar+ $^{12}$C,
$^{20}$Ne+ $^{28}$Si, and $^{24}$Mg+ $^{24}$Mg reactions at $E_{lab}$=187.7
MeV, 78.0 MeV, and 88.8 MeV, respectively. Spectra derived from experiment are
indicated by the thick-line histograms.  Spectra obtained from the
transition-state model calculation using the saddle-point method are indicated
by the thin-line histograms.  The vertical lines in (d) mark the mutual
excitations below 18 MeV involving the ground-state rotation band and all yrast
states, respectively. }
\label{fig35}
\end{figure}

\begin{figure}
\caption{ (a) Experimental and (b) calculated $\gamma$-ray spectra for the
$^{24}$Mg( $^{32}$S, $^{28}$Si)$^{28}$Si reaction at $E_{c.m.}$ with 7.6 MeV$
\leq E_x\leq $16.7 MeV. A smooth background has been subtracted from the
experimental yields for this comparison.  The figure is from ref[49].} 
\label{fig36}
\end{figure}

\begin{figure}
\caption{ Same as previous figure but with an expanded scale.  The figure is
from ref.[49]. } 
\label{fig37}
\end{figure}

\begin{figure}
\caption{ Comparison of the predicted mass distributions using the
transition-state model calculations (solid histograms) with the experimental
results (open histograms) for the $^{32}$S+ $^{24}$Mg reaction at $E_{lab}$=121
MeV and 142 MeV.  The figure is from ref.[45]. } 
\label{fig38}
\end{figure}

\begin{figure}
\caption{ Differential center-of-mass cross sections (points) as a function of
fragment mass for the (a) $^{40}$Ca+ $^{40}$Ca reaction at $E_{lab}$ =197 MeV
and $\theta_{lab}=20^o$ [121] and the (b) $^{28}$Si+ $^{50}$Ca reaction
at $E_{lab}$=150 MeV and $\theta_{lab}=30^o$ [122]. The figure is from
ref.[39]. } 
\label{fig39}
\end{figure}

\end{document}